\documentclass[12pt]{article}
\usepackage{amsmath}
\usepackage{amscd}
\usepackage{array}
\usepackage{latexsym,amssymb}

\newcommand{\nn}{\nonumber}

\def\sx{\left}
\def\dx{\right}
\def\to{\rightarrow}

\newcommand{\eq}[1]{(\ref{#1})}

\newcommand{\be}{\begin{equation}}
\newcommand{\ee}{\end{equation}}
\newcommand{\bea}{\begin{eqnarray}}
\newcommand{\eea}{\end{eqnarray}}

\newcommand{\ba}{\begin{eqnarray}}
\newcommand{\ea}{\end{eqnarray}}

\def\a{\alpha}

\def\b{\beta}

\def\d{\delta}
\def\e{\epsilon}

\def\g{\gamma}

\def\l{\lambda}

\def\o{\omega}

\def\s{\sigma}

\def\D{\Delta}
\def\G{\Gamma}
\def\L{\Lambda}
\def\O{\Omega}
\def\S{\Sigma}

\def\cL{\mathcal{L}}
\def\cM{\mathcal{M}}

\def\cQ{\mathcal{Q}}
\def\cR{\mathcal{R}}

\def\tI{{\Lambda}}
\def\tJ{{\Sigma}}
\def\tK{{\Gamma}}

\def\ii{\mathrm{i}}
\def\ib{{\ol {\imath}}}

\def\jb{{\ol {\jmath}}}

\def\mb{{\ol  m}}
\def\nb{{\ol  n}}
\def\rb{{\ol {r}}}
\def\sb{{\ol {s}}}

\def\notin{\hbox{{$\in$}\kern-.51em\hbox{/}}}

\def\inbar{\vrule height1.5ex width.4pt depth0pt}
\def\IB{\relax{\rm I\kern-.18em B}}
\def\IC{\relax\,\hbox{$\inbar\kern-.3em{\rm C}$}}
\def\ID{\relax{\rm I\kern-.18em D}}
\def\IE{\relax{\rm I\kern-.18em E}}
\def\IF{\relax{\rm I\kern-.18em F}}
\def\IG{\relax\,\hbox{$\inbar\kern-.3em{\rm G}$}}
\def\IH{\relax{\rm I\kern-.18em H}}
\def\II{\relax{\rm I\kern-.17em I}}
\def\IK{\relax{\rm I\kern-.18em K}}
\def\IL{\relax{\rm I\kern-.18em L}}
\def\IN{\relax{\rm I\kern-.18em N}}
\def\IP{\relax{\rm I\kern-.18em P}}
\def\IQ{\relax\,\hbox{$\inbar\kern-.3em{\rm Q}$}}
\def\IR{\relax{\rm I\kern-.18em R}}
\def\IU{\relax\,\hbox{$\inbar\kern-.3em{\rm U}$}}
\def\ZZ{\relax\ifmmode\mathchoice{\hbox{\cmss Z\kern-.4em Z}}{\hbox{\cmss Z\kern-.4em Z}}{\lower.9pt\hbox{\cmsss Z\kern-.4em Z}} {\lower1.2pt\hbox{\cmsss Z\kern-.4em Z}}\else{\cmss Z\kern-.4em Z}\fi}
\def\IGam{\relax{{\rm I}\kern-.18em \Gamma}}

\def\bfnull{\relax{\rm O \kern-.635em 0}}

\def\de{{\rm d}}

\def\na{\nabla}
\def\ol{\overline}

\def\imez{\frac{{\rm i}}{2}}
\def\mez{\frac{1}{2}}
\def\qu{\frac{1}{4}}

\def\square{{\,\lower0.9pt\vbox{\hrule \hbox{\vrule height 0.2 cm \hskip 0.2 cm \vrule height 0.2 cm}\hrule}\,}}

\def\twomat#1#2#3#4{\left(\begin{array}{cc} \end{array} \right)}

\begin{document}

\numberwithin{equation}{section}

\begin{center}
{\Large
$D=4$, $N=2$ Supergravity in the Presence of \\
Vector-Tensor Multiplets and the Role of higher $p$-forms
\\
\vskip 2mm
in the Framework of Free Differential Algebras }

\vskip 1.5cm

{\bf \large Laura Andrianopoli$^{1,2}$, Riccardo D'Auria$^2$\\ and
  Luca Sommovigo$^3$}\\
\vskip 8mm
 \end{center}
\noindent {\small{\it $^1$ Centro E. Fermi, Compendio Viminale,
    I-00184 Rome, Italy  and CERN-PH-TH, CH - 1211 Geneva 23;\\
    $^2$ Dipartimento di Fisica, Politecnico di Torino, Corso Duca
    degli Abruzzi 24, I-10129 Turin, Italy and Istituto Nazionale di
    Fisica Nucleare (INFN) Sezione di Torino, Italy\\ E-mail:  {\tt
      laura.andrianopoli@polito.it}; {\tt riccardo.dauria@polito.it}\\
    $^3$ Departament de F\'{\i}sica Te\`{o}rica, Universitat de
    Val\`{e}ncia and IFIC,\\ C/ Dr. Moliner, 50, E-46100 Burjassot
    (Val\`{e}ncia), Spain\\ E-mail: {\tt luca.sommovigo@uv.es}}}

\vfill

\begin{center}
{\bf Abstract}
\end{center}
{\small We thoroughly analyze at the bosonic level, in the framework
  of Free Differential Algebras (FDA), the role of 2-form potentials
  setting in particular evidence the precise geometric formulation of
  the anti-Higgs mechanism giving mass to the tensors. We then
  construct the (super)-FDA encoding the coupling of vector-tensor
  multiplets in $D=4$, $N=2$ supergravity, by solving the Bianchi
  identities in superspace and thus retrieving the full theory up to
  3-fermions terms in the supersymmetry transformation laws, leaving
  the explicit construction of the Lagrangian to future work. We
  further explore the extension of the bosonic FDA in the presence of
  higher $p$-form potentials, focussing our attention to the
  particular case $p=3$, which would occur  in the construction of
  $D=5$, $N=2$ supergravity where some of the scalars are properly
  dualized. }

\vfill\eject


\section{Introduction}
\label{intro}
The role of tensor multiplets in supergravity has seen in the last
years a revived interest, in connection with the study of flux
compactifications in superstring and M-theory. Indeed, it is well
known that in string theory one should expect $p$-forms of various
degree $p$ to enter in the non-perturbative formulation of the theory,
since they couple to extended objects ($p$ and $D$ branes).

The case of $p=2$ has been already discussed in the literature in
various contexts. Two-index antisymmetric tensors are 2-form gauge
fields whose field-strengths are invariant under the (tensor)-gauge
transformation $B \to B + \de \L$, $\Lambda$ being any 1-form. A
physical pattern to introduce massive tensor fields is the anti-Higgs
mechanism, where the dynamics allows the tensor  to take a mass  by a
suitable coupling to some vector field. The mass term plays the role
of magnetic charge in the theory. The investigation of the role of
massive tensor fields  was particularly fruitful for the $N=2$ theory
in 4 dimensions, where the study of the coupling of {\em
  scalar-tensor} multiplets (obtained by Hodge-dualizing  scalars
covered by derivatives in the hypermultiplet sector) in $N=2$
supergravity was considered, both as a CY compactification
\cite{Louis:2002ny} and at a purely four dimensional supergravity
level \cite{deWit:1982na,Theis:2003jj}. When this model was extended,
in \cite{Dall'Agata:2003yr,D'Auria:2004yi,Sommovigo:2004vj}, to
include the coupling to gauge multiplets, it allowed to construct new
gaugings containing also magnetic charges, and to find the
electric/magnetic duality completion of the $N=2$ scalar potential. As
far as the coupling of the {\it{vector-tensor multiplets}} is
concerned, the situation appears well established in five dimensional
supergravity, where a general formulation of the theory has been given
\cite{Gunaydin:1983rk,Gunaydin:1983bi,Gunaydin:1984pf,
Gunaydin:1984ak,Gunaydin:1984nt,Sierra:1985ax,Lukas:1998yy,
Lukas:1998tt,Gunaydin:1999zx,Gunaydin:2000xk,Gunaydin:2000ph,
Gunaydin:2003yx,Ceresole:2000jd,Bergshoeff:2004kh,Gunaydin:2005bf}.

The corresponding coupling of {\it{vector-tensor multiplets}} in
$N=2$, $D=4$ supergravity, which can be thought of as coming from
dualization of the real or imaginary part of the vector multiplet
scalars, can be studied by dimensional reduction from the $D=5,N=2$
theory \cite{Gunaydin:2005df,Gunaydin:2005bf}, thus catching important
properties of the couplings and of the gauge structure. However being
obtained by dimensional reduction this approach fails in giving the
most general four dimensional theory.

Another important step towards the construction of this theory has
been done in ref. \cite{deVroome:2007zd} by a thorough analysis
of the corresponding rigid supersymmetric theory in the framework of
the embedding tensor formalism.

A complete formulation is however  still missing, and in the present
paper we have filled this gap by working out the relevant superspace
Bianchi Identities coming from the most general $N=2$, $D=4$ Free
Differential Algebra (FDA) underlying the theory.

In the FDA approach a natural related issue is the coupling of higher
degree $p$-form potentials in supergravity theories as they can be
thought as non perturbative sectors of standard supergravities where
Hodge dualization of some field has been operated. However
supergravity formulations in $D$ dimensions coupled to  forms of order
higher than two have not been considered in detail until now, even if
some insight has been given for $p=3$ in the framework of maximally
extended supergravity \cite{deVroome:2007zd}. Indeed $p$-forms are
gauge fields $C^{(p)}$ whose field-strengths are invariant under the
(tensor)-gauge transformation $C^{(p)} \to C^{(p)} + \de \L^{(p-1)}$,
$\Lambda$ being any $(p-1)$-form. Being a gauge field, a $p$-form has
$\binom{D-2}{p}$ physical degrees of freedom, so that their existence
as propagating fields is limited to $p=2$ in four and to $p=3$ in five
dimensions, but higher $p$-forms can naturally appear in higher
dimensional supergravities. Note that their inclusion would give a
tool for the study of the non-perturbative structure of M or string
theory. \\We will show that, similarly to the $p=2$ case, higher rank
antisymmetric tensor fields may take mass via the Higgs mechanism
\footnote{The name of anti-Higgs mechanism was first introduced when
  the self-duality in odd dimensions for massive $2$-forms was
  discovered, since the degrees of freedom of the massive 2-form are
  obtained by absorbing those of a 1-form, which plays a passive role
  contrary to the active role played in ordinary Higgs mechanism where
  the vector absorbs a 0-form. However since the mechanism by which a
  $p$-form takes mass by absorbing the degrees of freedom of a
  $(p-1)$-form gauge field is, as we shall show, the same for any $p$,
  it should be called generalized Higgs mechanism, but we will adopt
  for simplicity the same locution "Higgs mechanism" as for the 0-form
  case.}.

In the present paper we study in  a general setting the structure of
the  FDA's since they give the natural framework for the  analysis of
theories including $p$-forms of any degree.

In the absence of supersymmetry, that is if we do not include the
gravitino 1-form, the general structure of the FDA is independent on
the space-time dimension $D$, the only restriction being that the
maximal degree of the $p$-form potentials is $D-1$. Instead, in the
supersymmetric case the FDA structure depends heavily on the
space-time dimension, according to the different properties of the
spinor representations of $SO(1,D-1)$. At the pure bosonic level the
general structure of the FDA with forms of any degree is very
complicate. However, if one restricts the degree of the forms to $p=1$
and $p=2$ the analysis can be done in full generality.

We therefore discuss in full detail first the case of $p=2$ both in the bosonic and in the supersymmetric
case. Before applying our consideration to $D=4,N=2$ supergravity in the presence of vector-tensor
multiplets we further study at the pure bosonic level the inclusion of $3$-forms in the FDA, leaving its
supersymmetric completion to a future investigation.

As far as 2-forms are concerned, we will first study at the bosonic
level and in full generality the algebraic structure which any theory
coupled to 2-index tensors and gauge vectors is based on. This
requires the extension of the notion of gauge algebra to that of FDA
that naturally accommodates in a general algebraic structure the
presence of $p$-forms ($p>1$) and gives a precise understanding of the
Higgs mechanism through which the antisymmetric tensors become
massive.  The discussion will be completely general, and will not rely
on the dimensions of space-time (apart from the obvious request $D\geq
4$, in order to have dynamical 2-forms) nor on supersymmetry.

The outcome of the analysis is that FDA approach allows to interpret
the resulting structure in a general group-theoretical way which is
not evident with other approaches. Even at the bosonic level, the
analysis of the FDA involving gauge vectors and $2$-forms charged
under a subalgebra of the gauge algebra, gives a geometric insight
into the structure of the physical theories it underlies. Indeed we
find that the inclusion of charged tensors implies a deformation of
the general gauge structure which can be precisely codified in terms
of a deformation of the structure constants and couplings of the gauge
group. Moreover, we obtain a precise algebraic understanding of the
Higgs mechanism through which the 2-forms can become massive.

Given the general structure of the bosonic case, we can then proceed to the supersymmetric extension of the
FDA underlying the $D=4$, $N=2$ supergravity theory. This can be done in two directions: either we analyze
the resulting $D=4$, $N=2$ theory in the presence of scalar-tensor multiplets, or in the presence of vector-tensor
multiplets. The former case has already been treated in ref. \cite{Dall'Agata:2003yr,D'Auria:2004yi}. As
already announced we focus our analysis on the latter theory, thus obtaining its general formulation. Indeed
the supersymmetrization of the FDA allows us to solve explicitly the Bianchi identities in superspace thus
obtaining the full theory, since, as it is well known, such solution implies the knowledge of the
supersymmetry transformation laws and the equations of motion.

The result of the analysis of the Bianchi identities shows the peculiar feature that a consistent
description of the FDA requires the simultaneous presence of fields related by Hodge duality, that is,
besides the electric potentials $A^M_\mu$ and  the antisymmetric tensors  $B_{M|\mu\nu}$, also their
Hodge-dual magnetic potentials $A_{M|\mu}$ and scalars $Y^M$. Actually, the fields $A_{M|\mu}$ and $Y^M$
will be recognised as the Hodge duals of $A^M_\mu$ and    $B_{M|\mu\nu}$ only after implementation of the
Bianchi identities, that is on shell.\footnote{This kind of approach was already pursued, at the lagrangian level and
for the maximally extended $D=4$, $N=8$ theory, in \cite{de Wit:2007mt}. } This peculiarity has two important
consequences. First of all, in absence of vector multiplets, the scalar potential of $N=2$ supergravity
coupled to vector-tensor multiplets is symplectic invariant. Secondly, the K\"ahler--Hodge structure of the
$\sigma$-model, which describes the off-shell geometry of the theory, can be and actually is different from
the on-shell geometry found after dualization. The study of the minima of the scalar potential requires the
knowledge of the on-shell $\sigma$-model geometry.
The construction of the Lagrangian in the rheonomic approach invariant under such transformations and giving
(in a simpler way) the equations of motion is left to future work.

The construction of the $N=2$ supersymmetric theory in five dimensions
can be done along the same lines, but since it reproduces the existing
results in the literature, we do not give its explicit construction
here.  We limit ourselves to make some remarks on the origin of the
self-duality of the tensors in terms of the Higgs mechanism giving
mass to 2-forms which were originally massless.

As far as the inclusion of $p$-forms with $p\geq3$ is concerned, we
have further analyzed in detail the FDA bosonic structure. Such
inclusion is particularly interesting  in five dimensional
supergravity for the case $p=3$, since a scalar is Hodge-dual to a
3-form.{\footnote{Note that $p=3$ can also play a role in $D=4$ since
    its field-strength can appear as a flux in the compactification
    from higher dimensions.}} In the standard $D=5$, $N=2$
supergravity, we have two possible kinds of scalars, namely the
scalars of the vector multiplets and the ones in the
hypermultiplets. When the vectors are dualized into 2-index tensors,
we further have scalars in the tensor multiplets. In each case, any of
these scalars can be dualized to 3-forms, originating new couplings
pertaining to different non-perturbative sectors of the theory. The
supersymmetrization of the relevant FDA in order to construct the
corresponding supergravity theory is under investigation.

Finally, considering higher $p$-forms, we also propose a possible
bosonic extension of the structure obtained for $p=3$ to any forms
with $p>3$, giving a completely consistent (bosonic) FDA.

The paper is organized as follows:

In section \ref{generalities} we study the general FDA describing the
coupling of two-index antisymmetric tensor fields to non-abelian gauge
vectors and show in detail, for the general case, how the Higgs
mechanism giving mass to the 2-forms takes place.

In section \ref{3fp} we include 3-forms in the FDA structure,
discussing different possible cases and give a possible extension of
the bosonic FDA including forms of any degree.

In section  \ref{susy4}, we apply the formalism to the case of $N=2$
four dimensional supergravity coupled to vector-tensor multiplets. In
our geometric approach the construction of the theory is obtained by
solving the Bianchi identities in superspace.
This allows us to  find the supersymmetry transformations rules, the
constraints on the scalar geometry which define the appropriate
$\s$-model of the theory, the fermionic shifts and the scalar
potential. As for the scalar-tensor theory we find that the
contribution from the vector-tensor sector to the scalar potential is
symplectic invariant.

In section \ref{susy5} we make some remarks on the issue of
self-duality in five dimensions for massive 2-forms.

The   Appendices contain technical details. \\
Appendix \ref{constraints3f} gives  the constraints arising in the
bosonic Bianchi identities for the FDA in the presence of 3-forms;\\
Appendix \ref{biss} describes the superspace solution of the Bianchi
identities for $D=4$, $N=2$ vector-tensor theory;\\
Appendix \ref{dualization} outlines the dualization procedure to
obtain the vector-tensor $\sigma$-model metric after dualization;\\
Appendix \ref{conv} contains our conventions and notations.


\section{A general bosonic  theory with massive 2-index tensors and
  non-abelian vectors}
\label{generalities}

In this section we are going to study the gauge structure of a general
theory with two-index antisymmetric tensor fields coupled to gauge
vectors. The discussion here will be general, with no need to make
reference to any particular dimension of space-time nor to any
possible supersymmetric extension of the model. Later, in section
\ref{susy4}, we will consider the supersymmetrization of the model,
specifying the discussion to the case of four dimensional  $N=2$
supergravity coupled to vector and vector-tensor multiplets.

\subsection{FDA and the anti-Higgs mechanism}
\subsubsection{Abelian case}
The simplest case of a FDA including 1-form and 2-form potentials
\footnote{0-forms will also be included in section \ref{susy4} when
  considering  supersymmetric versions of the theory} is described by
a set of abelian gauge vectors $A^M$ and of massless tensor
two-forms $B_M$ ($M=1,\dots n_T$.) interacting  by a coupling
$m^{MN}$. The field-strengths are:
\begin{equation}
\left\{
\begin{array}{rcl}
F^M &=& \de A^M + m^{MN} B_N \\
H_M &=& \de B_M
\end{array}
\right.
\end{equation}
and are invariant under the gauge transformations:
\begin{equation}
\left\{
\begin{array}{rcl}
\delta A^M &=& \de \Theta^M- m^{MN} \L_N \\
\delta B_M &=& \de \L_M \\
\end{array}
\right.
\end{equation}
with $\Theta^M$ parameters of infinitesimal U(1) gauge transformations
and $\L_M$ one-form parameters of infinitesimal tensor-gauge
transformations of the two-forms $B_M$. In this case the system
undergoes the Higgs mechanism, and it is possible to fix the
tensor-gauge so that:
\begin{equation}
\left\{
\begin{array}{rcccl}
A^M &\to & A'^M &=& - m^{MN} \ol  \L_N \\
B_M &\to & B'_M &=& B_M + \de \ol  \L_M; \\
\end{array}
\right.
\end{equation}
In this way the gauge vectors $A^M$ disappear from the spectrum
providing the degrees of freedom necessary for the tensors to acquire
a mass, since:
\begin{equation}
\left\{
\begin{array}{rcl}
F'^M &=&  m^{MN} B_N \\
H'_M &=& \de B_M.
\end{array}
\right. \label{antihiggs0}
\end{equation}

\subsubsection{Coupling to a non-abelian algebra}
The model outlined above may be generalized by including the  coupling
of this system to $n_V$ gauge vectors $A^X$ ($X = 1, \dots n_V$), with
gauge algebra $G_0$, if the index $M$ of the tensors $B_M$ and of the
abelian vectors $A^M$ runs over a representation of $G_0$. In this
case the FDA becomes \footnote{We will generally assume, here and in
  the following, that the tensor mass-matrix $m^{MN}$ is
  invertible. In case it has some 0-eigenvalues, we will restrict to
  the submatrix with non-vanishing rank. This is not a restrictive
  assumption, because any tensor corresponding to a zero-eigenvalue of
  $m$ may be dualized to a gauge vector and so included in the set of
  $\{A^X \}$.}:
\begin{equation}
\left\{
\begin{array}{rcl}
F^X &=& \de A^X + \mez f_{YZ}{}^X  A^Y \wedge A^Z \\
F^M &=& \de A^M - T_{X N}{}^M A^X \wedge A^N + m^{MN} B_N \\
&\equiv& D A^M + m^{MN} B_N \\
H_M &=& \de B_M + T_{X M}{}^N A^X \wedge B_N + d_{X NM} F^X \wedge
A^N \\
&\equiv& D B_M + d_{X NM}  F^X \wedge A^N
\end{array}
\right. \label{couplings}
\end{equation}
Setting $F^X=0$ the first equations give the Cartan-Maurer equations
of the Lie Algebra $G_0$ dual to the formulation in terms of the
generators $T_X$ with structure constants $f_{YZ}{}^X$. Here $T_{X
  M}{}^N $ and $d_{X MN}$ are suitable couplings. The closure of the
FDA (${\de}^2 A^X = {\de}^2 A^M = {\de}^2 B_M = 0$) gives the
following constraints:
\begin{align}
f_{[XY}{}^W f_{Z]W}{}^\O &= 0 \label{gaugealgebra1}  \\
T_{[X | M}{}^P T_{Y ] P}{}^N & = \frac 12  f_{X Y}{}^Z
T_{Z M}{}^N\label{gaugealgebra2}  \\
T_{X M}{}^N &= -d_{X MP }m^{NP} =d_{X PM}m^{PN}\label{gaugealgebra3}
\\
T_{X N}{}^M m^{NP} &= -T_{X N}{}^P m^{MN}\label{gaugealgebra4}  \\
T_{XM}{}^P d_{YNP} &+ T_{XN}{}^P d_{YPM} - f_{XY}{}^Z d_{ZNM}{} =
0. \label{gaugealgebra5}
\end{align}
Eqs. \eq{gaugealgebra1}, \eq{gaugealgebra2} show in particular that
the structure constants $f_{YZ}{}^X$ do indeed close the algebra $G_0$
and that $T_{X M}{}^N$ are generators of $G_0$  in the representation
spanned by the tensor fields. eqs. \eq{gaugealgebra3} and
\eq{gaugealgebra4} imply:
\begin{equation}
\begin{array}{rcl}
m^{MN} &=& \mp m^{NM} \\
d_{X MN} &=& \pm d_{X NM}.\\
\end{array}
\end{equation}
Note that \eq{gaugealgebra5} is a consistency condition that, when
multiplied by $m^{PQ}$, is equivalent to \eq{gaugealgebra2} (upon use
of \eq{gaugealgebra4}). From the physical point of view it simply
expresses the gauge covariance of the constants $d_{XMN}$, while from
the geometric point of view equation (\ref{gaugealgebra5}) means that
$d_{XMN}$ are are cocycles of the Lie Algebra Chevalley cohomology, in
the given representation labelled by the indices $MN$.

When \eq{gaugealgebra1} - \eq{gaugealgebra5} are satisfied, the
Bianchi identities read:
\begin{equation}
\left\{
\begin{array}{rcl}
\de F^X + f_{YZ}{}^X  A^Y \wedge F^Z &=& 0 \\
\de F^M - T_{X N}{}^M A^X \wedge F^N &=& m^{MN} H_N \\
\de H_M + T_{X M}{}^N A^X \wedge H_N &=& d_{X MN} F^N \wedge F^X. \\
\end{array}
\right. \label{bisimplified}
\end{equation}
To see how the  Higgs mechanism works in this more general case, let
us give the gauge  and tensor-gauge transformations  of the fields
(including the non-abelian transformations belonging to $G_0$, with
parameter $\e^X$). One obtains:
\begin{equation}
\left\{
\begin{array}{rcl}
\delta A^X &=& \de \e^X + f_{Y Z}{}^X  A^Y \e^Z \equiv D \e^X \\
\delta A^M &=& \de \Theta^M - T_{X N}{}^M A^X \Theta^N + T_{X
N}{}^M A^N \e^X - m^{MN} \L_N \\
&\equiv& D \Theta^M + T_{X N}{}^M A^N \e^X  - m^{MN} \L_N \\
\delta B_M &=& \de \L_M + T_{X M}{}^N A^X \wedge \L_N - d_{X NM}
F^X \Theta^N - T_{X M}{}^N B_N \e^X \\
&\equiv& D \L_M - d_{XNM} F^X \Theta^N - T_{X M}{}^N
B_N  \e^X ,\\
\end{array}
\right. \label{gaugeinvar1}
\end{equation}
with:
\begin{equation}
\left\{
\begin{array}{rcl}
\delta F^X &=& f_{Y Z} {}^X F^Y \e^Z \\
\delta F^M &=& T_{X N}{}^M F^N   \e^X \\
\delta H_M &=& - T_{X M}{}^N H_N \e^X.\\
\end{array}
\right.
\end{equation}
Fixing the gauge of the tensor-gauge transformation as:
\begin{equation}
\left\{
\begin{array}{rcl}
A^M &\to& A'^M = - m^{MN} \ol  \L_N \\
B_M &\to& B'_M = B_M + D \ol  \L_M,\\
\end{array}
\right. \label{gaugefix1}
\end{equation}
we find:
\begin{equation}
\left\{
\begin{array}{rcl}
F'^M &=& m^{MN} B_N \\
H'_M &=& D B_M \\
\end{array}
\right. \label{gaugefixed1}
\end{equation}
When the tensor-gauge is fixed as in \eq{gaugefix1},\eq{gaugefixed1},
the vectors $A^M$ disappear from the spectrum while the tensors $B_M$
acquire a mass. As anticipated in the introduction, this is in
particular the starting point of the  formulation adopted in the
literature to describe $D=5$, $N=2$ supergravity coupled to massive
tensor multiplets
\cite{Gunaydin:1999zx,Gunaydin:2000xk,Gunaydin:2000ph,Gunaydin:2003yx,Ceresole:2000jd,Bergshoeff:2004kh}.

\bigskip

However, let us observe that in this more general case the abelian
gauge vectors $A^M$, providing the degrees of freedom needed to give a
mass to the tensors via the anti-Higgs mechanism, are charged under
the gauge algebra $G_0$. It is not possible to make the gauge
transformation of the vectors $A^M$ compatible with that of the $A^X$
unless all together the vectors $\{A^X , A^M\} \equiv A^{\Lambda} $
form the co-adjoint representation of the larger non semisimple gauge
algebra $G= G_0 \ltimes \IR^{n_T}$.

The relations so far obtained may then be written with the collective
index ${\Lambda} =(X , M)$, in terms of  structure constants
$f_{{\Sigma}{\Gamma}}{}^{\Lambda}$ restricted to the following non
vanishing entries:
\begin{equation}
f_{{\Sigma}{\Gamma}}{}^{\Lambda}= ( f_{XY}{}^Z ,  f_{X
  M}{}^N\equiv -T_{X M}{}^N)\,, \label{algebrasimplified}
\end{equation}
and of the couplings:
\begin{equation}
m^{\Lambda M}\equiv \delta^{\Lambda}_N m^{NM}\,, \qquad d_{\Lambda
  \Sigma M} \equiv  \delta_{\Lambda}^X \delta_{\Sigma}^N d_{X
  NM}.\label{restrictions}
\end{equation}
When this notation is used the FDA \eq{couplings} reads:
\begin{equation}
\left\{
\begin{array}{rcl}
F^\L &\equiv& \de A^\L + \mez f_{\S\G}{}^\L A^\S \wedge A^\G + m^{\L
  M} B_M \\
H_M &\equiv& \de B_M + T_{\L M}{}^N A^\L \wedge B_N + d_{\L\S M} F^\L
  \wedge A^\S \\
\end{array}
\right. \label{fda}
\end{equation}
 with Bianchi identities:
\begin{equation}
\left\{
\begin{array}{rcl}
&\de F^\L& + \sx( f_{\S\G}{}^\L + m^{\L M} d_{\G\S M} \dx)A^\S \wedge
F^\G = \; m^{\L M} H_M \\
&\de H_M& +\sx( T_{\L M}{}^N + m^{\S N} d_{\S\L M} \dx) A^\L \wedge
H_N = \; d_{\L\S M} F^\L \wedge F^\S \\
\end{array}
\right.
\label{BI}
\end{equation}
provided the following relations hold:
\begin{equation}
\begin{array}{rcl}
f_{[\L\S}{}^\D f_{\G]\D}{}^\Pi &=& 0 \\
\sx[ T_\L , T_\S \dx] &=& f_{\L\S}{}^\G T_\G \\
T_{\L M}{}^{(N} m^{\L |P)} &=& 0 \\
m^{\L N} T_{\S N}{}^M &=& f_{\S\G}{}^\L m^{\G M} \\
T_{\L M}{}^N &=& d_{\L\S M} m^{\S N} \\
T_{[\L | M}{}^N d_{\G | \S] N} &-& (f_{[\L | \G}{}^\D  + m^{\D N}
  d_{\G [\L N} ) d_{\D | \S] M} - \mez f_{\L\S}{}^\D d_{\G\D  M} = 0.
\end{array}
\label{closure}
\end{equation}
Subject to the constraints \eq{closure}, the system is covariant under
the gauge transformations:
\begin{equation}
\left\{
\begin{array}{rcl}
\d A^\L &=& \de \e^\L + f_{\S\G}{}^\L A^\S \e^\G - m^{\L M} \L_M \\
\d B_M &=& \de \L_M + T_{\L M}{}^N A^\L \wedge \L_N - d_{\L\S M} F^\L
\e^\S - T_{\L M}{}^N \e^\L B_N \\
\end{array}
\right. \label{gaugefin}
\end{equation}
implying  the gauge transformation of the field strengths:
\begin{equation}
\left\{
\begin{array}{rcl}
\d F^\L &=& - \sx( f_{\S\G}{}^\L + m^{\L M} d_{\G\S M} \dx) \e^\S F^\G
\\
\d H_M &=& - \sx( T_{\L M}{}^N + m^{\S N} d_{\S\L M} \dx) \e^\L H_N \\
\end{array}
\right. \label{labfalfa}
\end{equation}

\subsubsection{Extending the FDA with a more general gauge group}
We now observe that the restrictions on the couplings
\eq{algebrasimplified} and \eq{restrictions} have been set to exactly
reproduce eqs. \eq{couplings} while exhibiting the fact that
$A^{\Lambda}$ collectively belong to the adjoint of some algebra
$G\supset G_0$. Actually eqs. \eq{couplings} and \eq{closure} allow in
fact a more general gauge structure than the one declared in
\eq{algebrasimplified}, \eq{restrictions}. Let $T_{\Lambda}\in {\rm
  Adj}\,G$ be the gauge generators dual to $A^{\Lambda}$. For the case
of \eq{algebrasimplified}, $G$ has the non-semisimple structure
$G=G_0\ltimes \IR^{n_T}$, and the generators $T_X \in G_0$ may be
realized in a  block-diagonal way (with entries $T_{X Y}{}^Z
=f_{XY}{}^Z$, $T_{X M}{}^N = - f_{X M}{}^N$) while the $T_M$ are
off-diagonal (with entries $T_{M X}{}^N =f_{X M}{}^N$). However, any
gauge algebra $G$ with structure constants $f_{\Lambda
  \Sigma}{}^{\Gamma}$ may in principle be considered, provided it
satisfies the constraints \eq{closure}. In the general case, to match
\eq{closure} one must also relax the restrictions on the couplings
\eq{algebrasimplified}, \eq{restrictions} and allow for more general
$f_{\Lambda \Sigma}{}^{\Gamma}$ and $d_{\Lambda\Sigma M}$. This
includes in particular the case
\begin{equation}
f_{XY}{}^M \neq 0\,, \qquad d_{XY M} \neq 0 \label{flsm}
\end{equation}
which was considered in \cite{Bergshoeff:2004kh} and
\cite{deWit:2004nw}. In this case, $G$ has not anymore in general a
non-semisimple structure, and $G_0$ is not a subalgebra of $G$
\footnote{We acknowledge an enlightening discussion with
  Maria~A.~Lled\'o on this point.}. This implies that the vectors
  $A^M$ do not decouple anymore at the level of gauge algebra, and
  this, at first sight, would be an  obstruction to implement the
  Higgs mechanism on the 2-forms. However, this apparent obstruction
  may be simply overcome in the FDA framework, due to the freedom of
  redefining the tensor fields as \cite{Dall'Agata:2005mj}:
\begin{equation}
B_M \to B_M  + k_{{\Lambda}{\Sigma} M} A^{\Lambda} \wedge A^{\Sigma}
,\label{red}
\end{equation}
for any $k_{{\Lambda}{\Sigma} M} $ antisymmetric in
${\Lambda},{\Sigma}$. It is then possible to implement the  Higgs
mechanism with the tensor-gauge fixing (which includes a field
redefinition as in \eq{red}):
\begin{equation}
\left\{
\begin{array}{rcl}
A^X &\to& A'^X =  A^X \\
A^M &\to& A'^M = - m^{ MN} \ol  \L_N \\
B_M &\to& B'_M = B_M  - \frac 12 d_{XY M}A^X \wedge A^Y + D \ol
\L_M \\
\end{array}
\right. \label{gaugefix}
\end{equation}
This still gives:
\begin{equation}
\left\{
\begin{array}{rcl}
F'^X &=& F^X \\
F'^M &=&  m^{MN} B_N \\
H'_M &=& D B_M \\
\end{array}
\right. \label{gaugefixed}
\end{equation}
provided that:
\begin{equation}
m^{MN} d_{[XY] N} = f_{XY}{}^M.
\end{equation}
With this observation, we may now  analyze in full generality which
non trivial structure constants may be turned on in \eq{fda} in a way
compatible with the anti-Higgs mechanism.

First of all, it is immediate to see that one must require:
\begin{equation}
f_{\Lambda M}{}^X = 0 \,,
\end{equation}
otherwise it is impossible to implement the Higgs mechanism. Indeed,
such structure constants  introduce a coupling to the gauge vectors
$A^M$  in the field-strengths $F^X$ which is not possible to reabsorb
by any field-redefinition.

Considering then the case:
\begin{equation}
f_{MN}{}^P \neq 0 \,, \qquad d_{MNP}\neq 0 \,.
\end{equation}
we see that $f_{MN}{}^P$ would introduce a non-abelian interactions
among the vectors $A^M$ and in particular, for the case $X =0$, this
would imply that the $A^M$ close a non-abelian gauge algebra. This
case may be treated in a way quite similar to the case \eq{flsm},
since again we may use the freedom  in \eq{red} to absorb the
non-abelian contribution to $F^M$ in a redefinition of $B_M$. The
anti-Higgs mechanism may then be implemented via the tensor-gauge
fixing:
\begin{equation}
\left\{
\begin{array}{rcl}
A^X &\to& A'^X =  A^X \\
A^M &\to& A'^M = - m^{ MN} \ol  \L_N \\
B_M &\to& B'_M = B_M  - \frac 12 d_{NP M}A^N \wedge A^P + D \ol
\L_M \\
\end{array}
\right. \label{gaugefix2}
\end{equation}
giving, as before:
\begin{equation}
\left\{
\begin{array}{rcl}
F'^X &=& F^X \\
F'^M &=&  m^{MN} B_N \\
H'_M &=& D B_M \\
\end{array}
\right. \label{gaugefixed2}
\end{equation}
provided that:
\begin{equation}
m^{MQ} d_{[NP] Q} = f_{NP}{}^M. \label{dmnp}
\end{equation}
This shows that also non-abelian gauge vectors $A^M$ may be
considered, and still may  decouple from the gauge-fixed theory by
giving mass to the tensors $B_M$. For this case, however, the
constraints \eq{closure}, together with \eq{dmnp}, give the following
conditions on the couplings:
\begin{equation}
\left\{
\begin{array}{rcl}
d_{MNP} &=& d_{[MNP]} \\
m^{MN}&=&+m^{NM} \\
\end{array}
\right. \label{msymm}
\end{equation}
Note that for the $D=5$, $N=2$ theory the matrix $m^{MN}$ turns out to
be antisymmetric, and this then implies, for that theory, $f_{MN}{}^P
=0$ \footnote{In this case, since any invertible antisymmetric matrix
  may be chosen as symplectic metric, then equation  $m^{M N}
  T_{{\Sigma} N}{}^P = - T_{{\Sigma}N}{}^M m^{N P}$ (from
  \eq{closure}) implies that the generators  $T_{\Lambda M}{}^N$
  belong to a symplectic representation of the gauge group
  \cite{Gunaydin:1984pf,Gunaydin:1984ak,Gunaydin:1984nt}.}.

\subsection{Some observations on the properties of the FDA}
\label{fdageneralities}
A further observation concerns eqs. \eq{BI} and \eq{labfalfa}. In
these equations, as in all the relations involving the physical field
strengths $F^{\Lambda}$ and $H_M$, the following objects appear:
\begin{equation}
\begin{array}{rcl}
\hat{f}_{{\Sigma}{\Gamma}}{}^{\Lambda} &\equiv&
f_{{\Sigma}{\Gamma}}{}^{\Lambda} + m^{{\Lambda} M} d_{{\Gamma}{\Sigma}
  M} \, ; \\
\hat{T}_{{\Lambda} M}{}^N &\equiv& T_{{\Lambda} M}{}^N + m^{{\Sigma}
  N} d_{{\Sigma}{\Lambda} M}= 2 d_{({\Lambda}{\Sigma})M} m^{{\Sigma}
  N}. \\
\end{array}
\label{generalgenerat}
\end{equation}

The generalized couplings $\hat{f}_{{\Sigma}{\Gamma}}{}^{\Lambda}$
belong to a representation of the gauge algebra $G$ which is not the
adjoint, since they are not antisymmetric in the lower indices. In
particular we find :
\begin{equation}
\begin{array}{rcl}
\hat f_{{\Lambda}{\Sigma}}{}^{\Gamma} m^{{\Sigma} M} &=& \hat
T_{{\Lambda} N}{}^M m^{{\Sigma} N} \\
\hat f_{{\Lambda}{\Sigma}}{}^{\Gamma} m^{{\Lambda} M} &=& 0. \\
\end{array}
\end{equation}
However, the $\hat{f}_{{\Sigma}{\Gamma}}{}^{\Lambda}$ and $\hat
T_{{\Lambda} N}{}^M$ can be understood as representations of
generators $\hat f_{\Lambda}$ and $\hat T_{\Lambda}$ that still
generate the gauge algebra $G$. Indeed the following relations hold
(subject to the constraints \eq{closure}):
\begin{equation}
\begin{array}{rcl}
\sx[ \hat{f}_{\Lambda} , \hat{f}_{\Sigma} \dx] &=& -
f_{{\Lambda}{\Sigma}}{}^{\Gamma} \hat{f}_{\Gamma},\\
\sx[ \hat{T}_{\Lambda} , \hat{T}_{\Sigma} \dx] &=&
f_{{\Lambda}{\Sigma}}{}^{\Gamma} \hat{T}_{\Gamma}.\\
\end{array}
\label{bigalgebra}
\end{equation}
The generalized couplings $\hat f$ and $\hat T$  express the
deformation of the gauge structure due to the presence of the tensor
fields. In particular, only the structure constants of $G_0$ are
unchanged, corresponding to the fact that this is the algebra realized
exactly in the interacting theory \eq{fda} after the anti-Higgs
mechanism has taken place. The rest of the gauge algebra $G$ is
instead spontaneously broken by the anti-Higgs mechanism (which
requires, if  $f_{XY}{}^M\neq 0$, also a tensor redefinition, as
explained in \eq{gaugefix}). However, the entire algebra $G$ is still
realized, even if in a more subtle way, as eqs. \eq{bigalgebra}
show. From a physical point of view, this is expected by a counting of
degrees of freedom, since the degrees of freedom required to make a
two-index tensor massive are the ones of a gauge vector connection
\footnote{Indeed, the on-shell degrees of freedom of a massless
  (2-index) tensor and of a vector in $D$ dimensions are
  $(D-2)(D-3)/2$ and $(D-2)$ respectively, while the ones of a massive
  tensor are $(D-1)(D-2)/2= (D-2)(D-3)/2 + (D-2)$.}, so that also the
vectors $A^M$, besides the $A^X$, are expected to be massless gauge
vectors. This algebra indeed closes provided the Jacobi identities
$f_{[{\Lambda}{\Sigma}}{}^{\Delta}  f_{{\Gamma}]{\Delta} }{}^{\Pi}
=0 $ are satisfied. We find indeed, using \eq{bigalgebra}:
\begin{equation}
\begin{array}{rcl}
\left[ \left[\hat f_{[{\Lambda}},\hat f_{\Sigma}\right],\hat
 f_{{\Gamma} ]} \right]_{\Delta}{}^{\Theta} &=& -
 f_{[{\Lambda}{\Sigma}}{}^{\Omega} f_{{\Gamma}]{\Omega} }{}^{\Pi} \hat
 f_{{\Pi} {\Delta} }{}^{\Theta} = 0 \\
\left[ \left[\hat T_{[{\Lambda}},\hat T_{\Sigma}\right], \hat
 T_{{\Gamma} ]} \right]_M{}^N &=& - f_{[{\Lambda}{\Sigma}}{}^{\Pi}
 f_{{\Gamma}]{\Pi}}{}^{\Delta} \hat T_{{\Delta}M}{}^N  =0 \\
\end{array}
\label{closed}
\end{equation}

The hatted generators $\hat f$, $\hat T$   play the role of {\em
  physical couplings} when the gauge structure is extended to include
charged tensors. They have then to be considered as the appropriate
generators of the free differential structure. It may be useful to
recast the theory in terms of all the couplings appearing in the
Bianchi identities \eq{BI}, that is the hatted generators and the
symmetric part $d_{({\Lambda}{\Sigma})M}$ of the Chern--Simons-like
coupling $d_{{\Lambda}{\Sigma} M}$. This is done by the field
redefinition:
\begin{equation}
B_M \to \tilde B_M = B_M + \frac 12 d_{[{\Lambda}{\Sigma}]M}
A^{\Lambda}\wedge A^{\Sigma} \label{bridef}
\end{equation}
so that the FDA takes the form:
\begin{equation}
\left\{
\begin{array}{rcl}
F^{\Lambda} &\equiv& \de A^{\Lambda} + \mez \hat
f_{{\Sigma}{\Gamma}}{}^{\Lambda} A^{\Sigma} \wedge A^{\Gamma} +
m^{{\Lambda} M}\tilde B_M \\
H_M &\equiv& \de \tilde B_M + \frac 12 \hat T_{{\Lambda} M}{}^N
A^{\Lambda} \wedge \tilde{B}_N + d_{({\Lambda}{\Sigma}) M}
F^{\Lambda} \wedge A^{\Sigma} + \\
&& + \mathcal{K}_{M {\Lambda}{\Sigma}{\Gamma}} A^{\Lambda} \wedge
A^{\Sigma} \wedge A^{\Gamma} \\
\end{array}
\right. \label{fda2}
\end{equation}
and the constraints \eq{closure} in the new formulation read, after
introducing $\tilde f_{{\Lambda}{\Sigma}}{}^{\Gamma} \equiv \hat
f_{[{\Lambda}{\Sigma}]}{}^{\Gamma}$:
\begin{equation}
\begin{array}{rcl}+
\tilde f_{[{\Lambda}{\Sigma}}{}^{\Pi} \tilde
  f_{{\Gamma}]{\Pi}}{}^{\Delta} &=& 2 m^{{\Delta} M}
  \mathcal{K}_{M[{\Lambda}{\Sigma}{\Gamma} ]} \\
\hat T_{[{\Lambda} M}{}^N  \hat T_{{\Sigma} ]N}{}^P  &=& \tilde
    f_{{\Lambda}{\Sigma}}{}^{\Gamma} \hat T_{\Gamma} + 12
    \mathcal{K}_{M{\Lambda}{\Sigma}{\Gamma}}m^{{\Gamma} P} \\
\hat T_{{\Lambda} M}{}^{N} m^{{\Lambda} P} &=& 0 \\
\frac{1}{2} m^{{\Lambda} N} \hat T_{{\Sigma} N}{}^M &=&\tilde
    f_{{\Sigma}{\Gamma}}{}^{\Lambda} m^{{\Gamma} M} \\
\hat T_{{\Lambda} M}{}^N &=& 2 d_{({\Lambda}{\Sigma}) M}
    m^{{\Sigma} N} \\
\hat T_{[{\Lambda} | M}{}^N d_{( {\Sigma}]{\Gamma} )N} &-& 2\hat
    f_{ [ {\Lambda} | {\Gamma}}{}^{\Delta} d_{ ({\Sigma}]{\Delta}) M}
  -\tilde f_{{\Lambda}{\Sigma}}{}^{\Delta}
    d_{({\Gamma}{\Delta}) M} = -6
  \mathcal{K}_{M{\Lambda}{\Sigma}{\Gamma}} \\
\mathcal{K}_{N [{\Sigma}{\Gamma}{\Delta}} \hat T_{{\Lambda}]|M}{}^N
  &-& 3  \mathcal{K}_{M {\Theta} [{\Lambda}{\Sigma}}\tilde
  f_{{\Gamma}{\Delta}]}{}^{\Theta} =0. \\
\end{array}
\label{closure2}
\end{equation}
In eqs. \eq{fda2} and \eq{closure2} we have introduced the definition:
\begin{equation}
\mathcal{K}_{M{\Lambda}{\Sigma}{\Gamma}}= \frac 12 \hat
f_{[{\Lambda}{\Sigma}}{}^{\Delta} d_{({\Gamma}] {\Delta})
  M} +\frac{2}{3} d_{({\Delta}[{\Sigma})M} \hat
  f_{{\Lambda}]{\Gamma}}{}^{\Delta}. \label{k}
\end{equation}
that could also be found by directly studying the closure of the FDA
\eq{fda2} without referring to its derivation from \eq{fda}.

Eq. \eq{fda2}, which is expressed in terms of the physical couplings
only,  is completely equivalent to \eq{fda}. This is in fact the
formulation used in \cite{deWit:2004nw}, for the study of $N=8$
supergravity in 5 dimensions.  However, as eqs. \eq{closure2} show, in
the formulation \eq{fda2} the gauge structure is not completely
manifest, because the Jacobi identities for the ``structure
constants'' $\tilde f_{\Lambda\Sigma}{}^{ \Gamma}$ fail to close.

Equation \eq{fda} (or, equivalently, \eq{fda2}) is the most general
FDA involving vectors and 2-index antisymmetric tensors.  Any other
possible deformation of \eq{fda} is indeed trivial.
\bigskip

As a final remark, let us observe that, given the definitions
\eq{fda}, the FDA still enjoys a  scale invariance under the
transformation, with parameter $\alpha$:
\begin{equation}
\left\{
\begin{array}{rcl}
m^{MN} &\to& \a \, m^{MN} \\
B_M &\to& \frac{1}{\a} B_M \\
d_{{\Lambda}{\Sigma} M} &\to& \frac{1}{\a} d_{{\Lambda}{\Sigma}
  M} \\
\end{array}
\right. \label{scaleinv}
\end{equation}
Such invariance is useful for fixing the overall normalization of  the
2-form contributions to the Chern--Simons terms when constructing the
Lagrangian.


\section{3-form potentials}
\label{3fp}
In this section we consider the extension of the bosonic FDA in the
presence of $p$-form potentials with $p\geq 3$. We focus our attention
in particular to 3-forms. Then,  in subsection \ref{genpot} we outline
a simple generalization of the bosonic FDA and of the related Higgs
mechanism providing a mass to higher degree $p$-form potentials.

The inclusion of 3-forms in a FDA has already been considered in
ref. \cite{deWit:2005hv} where its role in gauged maximal supergravity
was analyzed, with particular attention to the representation of the
duality group to which they belong.

Here, we consider the 3-forms as propagating fields with their
gauge-invariant field-strengths (4-form curvatures) appropriately
defined, that is  defined in such a way that the Bianchi identities
are satisfied in terms of gauge-invariant quantities. These
requirements are satisfied provided the various invariant tensors
entering the definition o the curvatures satisfy some constraints that
generalize those already found in the absence of 3-forms (see
eqs. \eq{closure} in section \ref{generalities}).

Note that if we think of  $p$-forms as coming from the dualization
of the scalar fields of some supergravity theory we can obtain new
theories where a subset of the original scalars are dualized into
$p$-forms. For example, if we consider   $D=5$, $N=2$ standard
supergravity, some scalar fields in the hypermultiplet sector could be
dualized into 3-forms, generating new couplings in the theory. By
dimensional reduction to $D=4$ of that theory, one should obtain the
$D=4$, $N=2$ supergravity coupled to scalar-tensor multiplets of
ref. \cite{Dall'Agata:2003yr,D'Auria:2004yi}. More general examples
should be worked out in the appropriate framework.

Since in our case we are interested in the general structure of the
bosonic FDA, the 3-form will be given a generic representation (lower)
index $\alpha$ and we shall denote it as $S_{\alpha}$. To construct
the new FDA we add $S_{\alpha}$ to the r.h.s. of the 3-form field
strength  $H_M$ and try to define a new 4-form field strength,
$G_{\alpha}$, in such a way that the Bianchi identities close. Let us
define the following FDA:
\begin{align}
F^\L &= \de A^\L + \mez f_{\S\G}{}^\L A^\S A^\G + m^{\L M} B_M \\
H_M &= \de B_M + T_{\L M}{}^N A^\L B_N + d_{\L\S M} F^\L A^\S +
L_{\L\S\G M} A^\L A^\S A^\G + k^\a_M S_\a \\
G_\a &= \de S_\a + T_{\L\a}{}^\b A^\L S_\b + T_{\a\L}{}^M F^\L B_M +
{T'}_{\a\L}{}^M A^\L H_M + \nn \\
& + L_{\L\S\G\a} A^\L A^\S F^\G + M_{\L\S\G\D\a} A^\L A^\S A^\G A^\D
+ N_{\a\L\S}{}^M A^\L A^\S B_M + \nn \\
& + N_\a{}^{MN} B_M B_N + R_{\a\L\S} F^\L F^\S \label{4form}
\end{align}
Here $k^\a_M$ is a matrix intertwining between the $M$- and the
$\alpha$- representations, and all the tensors appearing as
coefficients of the wedge products of forms in eqs. \eq{4form} are
constant tensors in the representations defined by the structure of
their indices, invariant under the action of the group $G$.

Differentiations of the curvatures $F^\L\, H_M, \, G_\a$ gives the
following Bianchi identities which are explicitly covariant under all
the local symmetries:
\begin{align}
D F^\L &= \de F^\L + \hat{f}_{\S\G}{}^\L A^\S F^\G = m^{\L M} H_M \\
D H_M &= \de H_M + T^{(H)}_{\L M}{}^N A^\L H_N = (d_{(\L\S)M} +
k_M^\a R_{\a\L\S}) F^\L F^\S + k^\a_M G_\a \\
D G_\a &= \de G_\a +  T^{(G)}_{\L\a}{}^\b A^\L G_\b = (T_{\L\a}{}^M +
{T'}_{\L\a}{}^M + 2 R_{\a\L\S} m^{\S M} ) F^\L H_M\,,
\end{align}
where we have defined:
\begin{align}
\hat{f}_{\S\G}{}^\L &= f_{\S\G}{}^\L + m^{\L M} d_{\G\S M} \\
T^{(H)}_{\L M}{}^N &= T_{\L M}{}^N + d_{\S\L M} m^{\S N} + k^\a_M
{T'}_{\L\a}{}^N \\
T^{(G)}_{\L\a}{}^\b &= T_{\L\a}{}^\b + {T'}_{\L\a}{}^M k_M^\b
\end{align}
provided that a set of  relations among the couplings are
satisfied. This is a complicated system of equations which may be
significantly simplified if we consider particular physical
situations. Since we will not develop further the general case here,
we  list the set of constraints in Appendix \ref{constraints3f}.
Among these identities, we just analyze the first equation in the
list, namely: $$m^{\L M} k^\a_M = 0 .$$ It is interesting to discuss
the physical meaning of such constraint. Roughly speaking, it
expresses the fact that, if we introduce in $H_M$ a non trivial
coupling $k^\a_M$ to give mass to the 3-form $S_\a$, then the mass
term $m^{\L M}$ for the corresponding 2-form $B_M$ must be zero. In
more detail, restricting our analysis to the case of $k^\a_M$ with
maximal rank, we will distinguish the possible different situations
which can arise depending on the presence or not of the coupling to
the  3-form potential, $k^\a_M$. We have then two substantially
different cases, corresponding to the range of the index $\a$ being
greater or lower to the range of $M$ (shortly $|\a | \geq |M|$ or
$|\a| < |M|$).

In the first case ($|\a | \geq |M|$), all the 3-form potentials may
become massive by absorbing the degrees of freedom of the 3-form
field-strengths $H_M$; this, however, can only be possible if the
d.o.f. of the $B_M$ potentials are those of massless fields, so that
the  Higgs mechanism giving mass to $B_M$ is actually forbidden. Let
us define, for this case, $S_M\equiv S_\a k^\a_M$, and call $S_a$ the
residual 3-form potentials (so that $S_\a \to (S_M,S_a)$). Then, by a
proper redefinition of the fields $S_\a$ and $B_M$, the FDA can always
be rewritten in the following form
\begin{align}
F^\L &= \de A^\L + \mez f_{\S\G}{}^\L A^\S A^\G \label{essealg1} \\
H_M &= \de B_M + T_{\L M}{}^N A^\L B_N  + S_M \label{essealg2} \\
G_M &= \de S_M + T_{\L M}{}^N A^\L S_N + T_{\L M}{}^N F^\L B_N +
R_{\L\S M} F^\L F^\S \label{essealg3} \\
G_a &= \de S_a
\end{align}
with Bianchi identities:
\begin{align}
D F^\L &= \de F^\L + f_{\S\G}{}^\L A^\S F^\G = 0 \label{bianchiS} \\
D H_M &= \de H_M + T_{\L M}{}^N A^\L H_N = \left(G_M -R_{\L\S M} F^\L
F^\S \right) \label{bianchis2} \\
D G_M &= \de G_M + T_{\L M}{}^N A^\L G_N = T_{\L M}{}^N F^\L H_N
\label{bianchis3} \\
\de G_a&= 0\,,\label{bianchis4}
\end{align}
where \begin{align}
\label{impr2}T_{\L M}{}^N &= k^\a_M T_{\L\a}{}^N \\
\label{impr8}T_{\L\a}{}^\b &= T_{\L\a}{}^M k_M^\b \,.
\end{align}
In the present case the set of constraints of Appendix
\ref{constraints3f} reduce dramatically to the following simple set of
constraints among the couplings :
\begin{align}
\label{impr1}f_{[\L\S}{}^\Pi f_{\G]\Pi}{}^\Delta &= 0 \\
\label{impr9}T_{[\L|M}{}^P T_{\S]P}{}^N &= \mez f_{\L\S}{}^\G T_{\G M}{}^N \\
\label{impr7}T_{\L M}{}^N k_N^\a &= k_N^\b T_{\L\b}{}^\a \\
T_{(\L|M}{}^N R_{\S)\G N} -2 f_{\L(\S}{}^\D R_{\G)\D M}&= 0\label{impr11}
\end{align}
One can verify that a further differentiation of the Bianchi
identities \eq{bianchiS} gives identically zero.

On the other hand, if $|\a |< |M|$, so that $M=(\a , \tilde M)$, a
subset of 2-index potentials, $B_{\tilde
M}$, does not disappear from the spectrum. In this case the usual
Higgs mechanism discussed in section 2
may take place for the 2-forms $B_{\tilde M}$, so that all the
relations involving 2- and 3-form
field-strengths are still valid; however, now the explicit solution of
the constraints \eq{mk} -
\eq{constrlast}   is more involved, and we leave it to forthcoming
work, where the appropriate applications
to five dimensional supergravity will be discussed.

Note that, in ref. \cite{deWit:2005hv}, the 3-form potentials were introduced on the r.h.s. of $H_M$ in the form
$\hat{T}_{\L M}{}^N S_N{}^\L $, where $\L$ is an index in the duality group of a supergravity theory. We
have seen that if we want the 3-form potentials as propagating physical fields in a supergravity Lagrangian
where the scalar fields of some supermultiplet have been dualized, then $\hat{T}_{\L M}{}^N$ must be
substituted with ${T}_{\L M}{}^N$, since $m^{\L M}=0$.

The construction of a supergravity theory of this kind, which amounts to extend the FDA to a super-FDA
containing the gravitino spinor 1-forms, particularly in the case of an $N=2$ five dimensional supergravity,
will be left to a future investigation. However, we may observe that there is no obstruction in enlarging
the FDA to a super-FDA, at least in five dimensions. Indeed, one can modify the definition of the curvatures
in superspace the gravitino 1-forms $\psi_A$, in such a way that the closure of the FDA is still achieved.
 Let us consider, as an example, the case where some of the scalars in the
hypermultiplet sector have been dualized into $S_\alpha$. As far as the 3-forms in superspace are concerned,
the proper definition of their curvatures is the following:
\begin{align}
G_\a &= \de S_\a + T_{\L\a}{}^\b A^\L S_\b + T_{\L\a}{}^M \left(F^\L -
X^\L \ol  \psi_A \psi^A\right) B_M + \nn \\
&+ L_{\L\S \a} \left(F^\L - X^\L \ol  \psi_A \psi^A\right) \left(F^\S
- X^\S \ol  \psi_A \psi^A\right) + \omega_\alpha^{AB} \ol  \psi_A
\G_{ab}\psi_B V^a V^b\,.
\end{align}
One immediately sees that the super-FDA at zero curvatures (minimal
FDA) is consistent owing to the five
dimensional Fierz identities
\begin{equation}
\ol \psi_C \G^{ab}\psi_A \ol \psi_B \G_{b}\psi^B = 0\,,
\end{equation}
provided $\nabla\omega_\alpha^{AB}=0$. Analogously to the case
analyzed  in \cite{Dall'Agata:2003yr},
$\omega_\alpha^{AB}$ has an interpretation as the $SU(2)$ connection
whose coordinate index is
Hodge-dualized into a 3-form tensor index $\alpha$. In fact we expect
that the dimensional reduction of this
theory would reproduce the $D=4$ $N=2$ supergravity coupled to
scalar-tensor multiplets given in
\cite{Dall'Agata:2003yr,D'Auria:2004yi}.
\subsection{A possible generalization to $p$-form potentials} \label{genpot}
 The analysis of the previous sections admits a generalization
  allowing the presence of $p$-form potentials,
  with $p >3$. Indeed, a possible generalization of the Higgs
mechanism, generating massive $p$-forms, can be achieved introducing
  the following set of curvatures:
\begin{align}
F^\L &= \de A^\L + \mez f_{\S\G}{}^\L A^\S A^\G \\
F^M &= \de A^M - T_{\L N}{}^M A^\L A^N + m^{MN} B_N \\
H^{(3)}_M &= \de B^{(2)}_M + T_{\L M}{}^N A^\L B^{(2)}_N + d_{\L NM}
F^\L A^N \\
H^{(3)}_\a &= \de B^{(2)}_\a + T_{\L\a}{}^\b A^\L B^{(2)}_\b +
S^{(3)}_\a \\
G^{(4)}_\a &= \de S^{(3)}_\a + T_{\L\a}{}^\b A^\L S^{(3)}_\b +
T_{\L\a}{}^\b F^\L B^{(2)}_\b \\
...&=............................ \nn\\
H^{(n-1)}_\a &= \de B^{(n-2)}_\a + T_{\L\a}{}^\b A^\L B^{(n-2)}_\b +
S^{(n-1)}_\a \\
G^{(n)}_\a &= \de S^{(n-1)}_\a + T_{\L\a}{}^\b A^\L S^{(n-1)}_\b +
T_{\L\a}{}^\b F^\L B^{(n-2)}_\b
\end{align}
Differentiating the curvatures, we get the following set of covariant
Bianchi identities:
\begin{align}
D F^\L &= \de F^\L + f_{\S\G}{}^\L A^\S F^\G = 0 \displaybreak[0] \\
D F^M &= \de F^M - T_{\L N}{}^M A^\L F^N = m^{MN} H_N \displaybreak[0] \\
D H_M &= \de H_M + T_{\L M}{}^N A^\L H_N = d_{\L NM} F^\L F^N \displaybreak[0] \\
D H_\a &= \de H_\a + T_{\L\a}{}^\b A^\L H_\b = G_\a \displaybreak[0] \\
D G_\a &= \de G_\a + T_{\L\a}{}^\b A^\L G_\b = T_{\L\a}{}^\b F^\L
H_\b \\
...&=............................ \nn\\
D H^{(n-1)}_\a &= \de H^{(n-1)}_\a + T_{\L\a}{}^\b A^\L H^{(n-1)}_\b
= G^{(n)}_\a \displaybreak[0] \\
D G^{(n)}_\a &= \de G^{(n)}_\a + T_{\L\a}{}^\b A^\L G^{(n)}_\b = T_{\L\a}{}^\b F^\L H^{(n-1)}_\b
\end{align}
provided that the following relations hold:
\begin{align}
f_{[\L\S}{}^\D f_{\G]\D}{}^\Pi &= 0 \\
\left[ T_\L , T_\S \right] &= f_{\L\S}{}^\G T_\G \label{algebraT} \\
T_{\L P}{}^M m^{PN} + m^{MP} T_{\L P}{}^N &= 0 \label{Tsimm1} \\
T_{\L N}{}^M &= - d_{\L NP} m^{MP} \label{Tsimm2} \\
T_{\L N}{}^M &= d_{\L PN} m^{PM} \label{Tsimm3} \\
T_{\L M}{}^P d_{\S NP} + d_{\S PM} T_{\L N}{}^P &= f_{\L\S}{}^\G d_{\G
  NM} \,.\label{algebraT2}
\end{align}

\section{$N=2$, $D=4$  Supergravity}
\label{susy4}

Our most important application of the previous formalism, restricted to 2-form potentials (see Section
\ref{generalities}), is the explicit construction of the $N=2$, $D=4$ supergravity theory coupled to
vector-tensor multiplets, that is those multiplets that can be obtained from vector multiplets by
Hodge-dualization of, say, the imaginary part of a subset of the complex scalar fields parametrizing the
special manifold. At our knowledge, as anticipated in the introduction, the construction of such theory in
full generality has not been achieved so far, even if important steps in that direction have been given in
ref. \cite{Gunaydin:2005bf}, where the four dimensional theory was obtained by dimensional reduction from
five dimensions and the ensuing properties thoroughly analyzed. However, this approach does not catch the
most general theory,  being restricted to models with a five dimensional uplift. A general approach
containing vector-tensor multiplets has been developed in ref. \cite{deVroome:2007zd} by use the framework
of the embedding tensor, but it is  restricted to supersymmetric rigid gauge theories.

The general analysis we develop in this section relies on the solution of Bianchi identities in superspace
which, besides giving the general supersymmetry transformation laws and the constraints on the geometry of
the relevant $\sigma$-models, also allows us in principle to retrieve the equations of motion of the theory.
It would be of course desirable to have the supersymmetric Lagrangian to put in evidence the couplings of
the theory, but we leave this construction to a forthcoming paper. However, the knowledge of the  explicit
expression of the fermion shifts will allow us to reconstruct the scalar potential.

 Let us consider $N=2$ supergravity in four dimensions with field content given by:
$$(V^a_\mu, \psi_{A\mu}, \psi^A_\mu, A^0_\mu)\,, $$
(where $a$ and $\mu$ denote space-time indices respectively flat and curved, $A=1,2$ is an $SU(2)$ index and
we have decomposed the gravitino in chiral ($\psi_A$) and anti-chiral ($\psi^A$) components), coupled to
$n_V$ vector multiplets:
$$(A_\mu, \lambda^{A}, z)^r\,,\qquad r=1,\cdots,n_V\,,$$
where $z^r$ are holomorphic coordinates on the special manifold
$\cM_V$ spanned by its scalar sector and $\lambda^{rA}$ are chiral
spin-1/2 fields (with complex conjugate antichiral component
$\lambda^\rb_A$), and to $n_T$ vector-tensor multiplets:
$$(B_{M\mu\nu},A^{M}_{\mu}, \chi^{m A}, \phi^m)\,,\qquad M,m=1,\cdots,n_T\,,$$ where $\phi^m$ are
 real coordinates on the real manifold $\cM_T$ spanned by its scalar sector
  ($m$ is a coordinate index on $\cM_T$ while $M$ is a representation index
   of the non-semisimple gauge group $G$) and $\chi^{m A}$ are Majorana spinors, not decomposed in chiral components.
   Written in this way, the tensor
   multiplets are naturally  interpreted as obtained  from $n_T$ extra vector multiplets
   by Hodge-dualization of the imaginary part $Y^m$ of the corresponding holomorphic scalars $z^m$, as will be apparent in the
   following (see eq. \eq{starH} and Appendix \ref{dualization}).

Let us now study here the supersymmetric FDA of this theory. We shall let all the vectors $A^{\L}$ be the
gauge vectors of a non abelian non-semisimple algebra $G=G_0\ltimes \IR^M$ and the tensors $B_M$ to be in a
representation of $G_0$.\footnote{As discussed in section 2, even if we start from a more general $G\supset
G_0 \ltimes \IR^{n_T}$, we can always retrieve this case by a suitable redefinition of the 2-forms $B_M$.}
 In the interacting theory, the anti-Higgs mechanism will take place so that the
vectors $A^M$ will provide the degrees of freedom to give mass to
the tensors $B_M$. In this way the gauge algebra will be broken to
its subalgebra $G_0$ ($\dim G_0= n_V+1$) spanned  by the vectors
$(A^0, A^r)$.  It will then be useful to adopt a collective
gauge-vector index $\L=(X,M)=0,1,\cdots,n_V +n_T$ (with $X
=0,1,\cdots,n_V$) running over all the vectors of the theory.

In order for the FDA to close at the supersymmetric level, it will be necessary to include among the
defining bosonic fields of the tensor multiplet sector, besides the vectors $A^M$ and the tensors $B_M$,
also their Hodge duals $A_M$ (${}^*dA^M \propto  m^{MN} dA_N$) and $Y^m$. The dual gauge vectors $A_M$
(undergoing a dual Higgs mechanism, since they take mass by eating the degrees of freedom of the dualized
scalars $Y^m$) will indeed appear in the supercurvature of the tensor field-strengths $H_M$. If we would not
include them, the Bianchi identities would show up inconsistencies. Then, since the fields $Y^m$ have to be included for a
correct description of the dynamics of the theory, it is convenient to adopt a complex notation also for the
vector-tensor sector and work with holomorphic coordinates $z^m \equiv \phi^m +\ii Y^m$ together with their
complex conjugates $\ol z^\mb \equiv \phi^m -\ii Y^m$.  Correspondingly, we will generally decompose the
Majorana spinor $ \chi^{mA}$ in chiral components denoted by $\lambda^{mA}, \lambda^\mb_A$. Using this
notation, it will be natural to introduce a collective holomorphic world-index $i=(r,m)=1,\cdots , n_V+n_T$,
in parallel to what has been done for gauge indices. At this point we note that using the collective index
formalism the theory looks much like the standard $N=2$ supergravity coupled to vector multiplets only, and
this explains, as we will see in the following, that most of the results coming from Bianchi Identities will
look formally like those of the standard $N=2$ supergravity. Since then the Free Differential Algebra
involves both the antisymmetric tensors $B_M$ and the scalars $Y^m$, we expect that the closure of the
Bianchi identities would imply the duality relation between them. Actually, this is what happens implying
that the dualization relation is valid only on-shell. As a consequence, the on-shell geometry will look
quite different from its off-shell (K\"ahler--Hodge) counterpart. In particular, in absence of a
factorization of the two $\sigma$-models, the off-shell K\"ahler--Hodge structure is completely destroyed since the
metric is not even hermitian.

We further note that, exactly like the five dimensional case, for the $N=2$ four dimensional theory the
massive vector-tensor multiplets are short, BPS multiplets.\footnote{This has to be contrasted to what
happens for the scalar-tensor multiplets where the tensor field is Hodge-dual to a scalar in the
hypermultiplet sector. In that case, the multiplet becomes massive by introducing an appropriate coupling to
a vector multiplet. In our case, instead, the fields $A_M$ and $Y^m$ do not have a spinor partner, but act
as bosonic Lagrange multipliers in the theory.} They are therefore charged and this in turn requires for CPT
invariance that the vector-tensor multiplet sector always includes an even number of tensor fields. We have
chosen to take as vectors belonging to the vector-tensor multiplets the vectors  $A^M$ with an upper
representation index, since these are the degrees of freedom participating to the  anti-Higgs mechanism
giving mass to the tensors $B_M$. As a byproduct, the Hodge-dual vectors $A_M$ become massive by eating the
degrees of freedom of the scalars $Y^m$ (Hodge-dual to the tensors $B_M$).

Passing from the bosonic FDA to the one for $N=2$ supergravity in four dimensions, the relations defining
the algebra get modified in various directions. First of all one has to include the FDA of pure
supergravity:
\begin{align}
\cR^a{}_b &= \de \o^a{}_b - \o^a{}_c \wedge \o^c{}_b
\label{curvature} \\
\mathcal{T}^a &= \de V^a - \o^a{}_b V^b - \ii \ol \psi_A \g^a \psi^A
\label{tors} \\
\rho_A &= \de \psi_A - \qu \o_{ab} \g^{ab} \psi_A + \imez \mathcal{Q}
\psi_A  \label{rhol} \\
\rho^A &= \de \psi^A - \qu \o_{ab} \g^{ab} \psi^A - \imez \mathcal{Q} \psi^A \label{rhor} \,.
\end{align}
where with $\mathcal{Q}\equiv Q_r dz^r + Q_{\bar r} d\ol z^{\bar r}$
we denote the gauged $U(1)$-K\"ahler
 composite connection of special geometry. \footnote{For its definition in terms of the ungauged one and for
  all the notation concerning special geometry, we refer
the reader to the standard $N=2$, $D=4$ supergravity of ref. \cite{Andrianopoli:1996cm}.}

 Secondly, the bosonic curvatures introduced in section \ref{generalities} for the field-strengths
have to be generalized to their supersymmetric extension, that is:
\begin{align}
H_M &= \de B_M + T_{\L M}{}^N A^{\tI} B_N  + \ii P_M \ol \psi_A \g_a
\psi^A V^a + \nn \\
&+ \left( d_{\tI\tJ M} A^\tJ + \hat T_{\tI M}{}^N A_N \right) \left(
F^\tI - L^\tI \ol \psi^A \psi^B \e_{AB} - \ol  L^\tI \ol \psi_A
\psi_B \e^{AB} \right) \label{HM} \\
F^\tI &= \de A^\tI + \mez f_{\tJ\tK}{}^\tI A^\tJ A^\tK + m^{\tI M} B_M
+ L^\tI \ol \psi^A \psi^B \e_{AB} + \ol  L^\tI \ol \psi_A \psi_B
\e^{AB} \label{FtI} \\
F_M &= \de A_M + \hat T_{\tI M}{}^N A^\tI A_N + M_M \ol \psi^A \psi^B
\e_{AB} + \ol M_M \ol \psi_A \psi_B \e^{AB} \label{FM}
\end{align}
Here $P_M$ is a real section on the $\sigma$-model, $L^\L$ and $\ol
L^\L$ are the sections of special geometry,  while $M_M$ and $\ol
M_M$ are new sections in the given representation of $G_0$.

 Finally, the FDA has to be enlarged to include the 1-form gauged
 field-strengths for
the 0-form scalars and spinors belonging to the representations of
 supersymmetry:
\begin{align}
D z^i &= \de z^i + k^i_\L A^\L  - k^{iM}A_M \label{zi} \\
\na \l^{iA} &= \de \l^{iA} - \qu \o_{ab} \g^{ab} \l^{iA} +
  \G^i{}_j \l^{jA}  \label{lambdai}
\end{align}
where $k^i_\L$ are the holomorphic part of Killing vectors in the adjoint representation of the algebra $G$
while $k^{iM}\equiv k^{mN}\delta_m^i$ are  purely imaginary Killing vectors in the coadjoint representation
of the invariant subgroup of $G$, satisfying $k^{mM} = - k^{\mb M}$. This choice corresponds to the
requirement that the vectors $A_M$ undergo the Higgs mechanism by eating the imaginary part $Y^m$ of the
scalars $z^m$. This will be consistent with the solution of the Bianchi identities.

The (on-shell) solution of the Bianchi identities  for the value of the supercurvatures, up to three-fermion
terms, is given in detail in Appendix \ref{param4d}. In the solution new structures appear, which are
defined by a set of constraints. On the scalar geometry one finds:
\begin{align}
D_i L^\L &= f^\L_i - \ii\, Q_m \, \delta_i^m \,L^\L\,, & D_{\ib} L^\L &=- \ii\,
 Q_\mb \,\delta_\ib^\mb \, L^\L \label{gli}\displaybreak[0] \\
D_i M_M  &= g_{Mi}- \ii\, Q_m \, \delta_i^m \,M_M  \,,& D_{\ib} M_M &=- \ii\, Q_\mb \,\delta_\ib^\mb \, M_M
\label{gmi} \displaybreak[0]
\end{align}
\begin{align}
h_a &= \frac \ii 2 \left(Q_m D_a z^m + Q_\mb D_a \ol z^\mb \right)  \label{ha} \\
P_M &= - \sx[ 2 d_{(\tI\tJ) M} L^\tI \ol  L^\tJ + \hat{T}_{\tI M}{}^N
  \sx( \ol  L^\tI M_N + L^\tI \ol  M_N \dx) \dx] \\
0 &= d_{(\tI\tJ) M} L^\tI L^\tJ + \hat{T}_{\tI M}{}^N L^\tI M_N
\label{constraint} \\
0 &= 2 d_{(\tI\tJ) M} L^\tI f_i^\tJ + \hat{T}_{\tI M}{}^N \sx( L^\tI
g_{iN}+f_i^\tI M_N \dx)  \\
D_i P_M &=  2h_{Mi} \label{hmi}
\end{align}
The new quantities $g_{Mi}$, $h_{Mi}$  are defined by eqs. \eq{gmi}, \eq{hmi}. Moreover, $Q_m, Q_\mb$ are
vectors on the $\sigma$-model of the vector-tensor scalars which can be thought as coming from the K\"ahler
connection of special geometry after dualization.

 As far as the field strengths appearing in the
fermions transformation laws are concerned, we find the following relations:{\footnote{We recall that in the
superspace rheonomic approach the components of the curvatures along the bosonic vielbein $V^a$ do not
coincide with their space-time components along the differentials $dx^\mu$. Actually, they differ from the
space-time components by fermion bilinears and they coincide, in the component approach, with the
supercovariant field strengths. The fermion bilinears are immediately retrieved from the superspace
parametrizations given in Appendix \ref{param4d} by projecting the supercurvatures $H_{Mabc}$,
$F^\Lambda_{ab}$, $F_{Mab}$, $Z^i_a$,  on the space-time differentials (see, for example,  Appendix A of
ref. \cite{Andrianopoli:1996cm}.).}}

\begin{align}
F^\L_{ab} &= 2 \sx( f^\L_i G^{i-}_{ab} + \ol f^\L_{\ib} G^{\ib
  +}_{ab} \dx) + \ii \sx( L^\L T_{ab}^+ + \ol L^\L T_{ab}^- \dx)  \label{FlambdaGT}\\
F_{M|ab} &= 2 \sx( g_{Mi} G^{i-}_{ab} + \ol g_{M\ib} G^{\ib +}_{ab}
  \dx)+ \ii \sx( M_M T^+_{ab} + \ol M_M T^-_{ab} \dx)
 \label{FMGT}\\
h_{Mi} G^{i-}_{ab} &= 0 \label{hG}\\
H_{Mabc} &=-\frac \ii 3 \e_{abcd} \sx[  \sx(  h_{Mi}
  D^d z^{i}-   h_{M \ib} D^d\ol  z^{\ib}  \dx) \dx] \label{starH}\,.
\end{align}

 Finally, on the gauge sector we obtain a set of relations involving the Killing vectors and the fermionic shifts.
 They can be split into an $SU(2)$-singlet sector and an $SU(2)$-adjoint sector, corresponding to the $U(1)$
 and $SU(2)$ parts of the R-symmetry.  Indeed, as we will see in the discussion of the scalar potential, the
 $U(1)$ part is related to the vector and vector-tensor couplings, while the $SU(2)$-part pertains to the
 hypermultiplets and scalar-tensor multiplets.

\paragraph{$U(1)$-sector relations:}

\begin{align}
&k^i_\L L^\L - \, k^{iN} M_N=0
  \,;\label{km}\displaybreak[0] \\
&W^{i [AB]} = \epsilon^{AB}\, \left(k^i_\L \ol  L^\L -k^{iM} \ol M_M\right) \,;\label{wiab} \displaybreak[0] \\
& Q_m W^{m[AB]} = Q_{\mb} \ol  W^{\mb}_{[AB]} = 0\,;\displaybreak[0] \\
& g_{Mi} \e_{AB} W^{iAB} + \ol g_{M\ib} \e^{AB}\ol W^\ib_{AB} =0\,; \displaybreak[0] \\
&P_M \,m^{\L M}  = - \mez f^\L_i \e_{AB} W^{iAB} - \mez \ol f^\L_\ib \e^{AB} \ol W^\ib_{AB} \,.
\end{align}
We note that eq. \eq{km} differs from the analogous one for special geometry by the presence on the
right-hand side of the magnetic Killing vector $k^{Mn}$. Furthermore, as eq. \eq{wiab} shows, the singlet
shift of the spinors in the vector-tensor multiplets is symplectic invariant, once the gauging of the vector
multiplets and of the graviphoton is turned off.

\paragraph{$SU(2)$-sector relations:}

\begin{align}
&P_M S^{AB}= h_{Mi} W^{i(AB)} \,;\\
& \e_{AC} \sx( f^{\L}_i W^{i (CB)} - 2 L^{\L}\ol S^{CB} \dx) +  \e^{BC} \sx( \ol{f}^{\L}_{\ib}\ol
W^{\ib}_{(AC)} - 2 \ol{L}^{\L} S_{AC}   \dx) = 0\,;\\
& \e_{AC} \sx( g_{Mi} W^{i (CB)} - 2 M_M \ol S^{CB}  \dx) + \e^{BC} \sx( \ol{g}_{M\ib} \ol W^{\ib}_{(AC)} -
2 \ol{M}_M S_{AC} \dx) = 0  \,. \label{constrgaugelast}
\end{align}
We observe that these relations are simple extensions of  those obtained for standard $D=4$, $N=2$
supergravity.

\bigskip

From the physical point of view the main interest  is of course in the supersymmetry transformation laws,
which are an immediate consequence of the solution for the curvatures given in Appendix \ref{param4d}. Up to
3-fermions\footnote{In eqs. \eq{trasf:rhol} and \eq{trasf:rhor} we have kept the 3-fermions terms of type
$\epsilon \psi\chi$ since they are essential in the analysis of the gauge fermion shifts.}
 they read:
\begin{align}
\delta V^a_\mu  &= -\ii \ol  \psi_{A\mu} \g^a \epsilon^A -\ii \ol
\psi^A_{\mu} \g^a \epsilon_A \label{trasf:tors} \displaybreak[0] \\
\delta B_{M\mu\nu} &=   h_{Mi} \ol \epsilon_A \g_{\mu\nu} \l^{iA}   +
\ol  h_{M\ib} \ol \epsilon^A \g_{\mu\nu} \l^{\ib}{}_A -\ii P_M
\left(\ol\epsilon_A \g_{[\mu}\psi_{\nu]}^A -\ol\psi_{A[\nu} \g_{\mu]}
 \epsilon^A\right)+ \nonumber\\
 & \qquad + 2\left( d_{\tI\tJ M} A^\tJ_{[\mu} + \hat T_{\tI M}{}^N
   A_{N[\mu} \right) \left(  L^\tI \ol \epsilon^A \psi_{\nu]}^B
\e_{AB} + \ol  L^\tI \ol \epsilon_A \psi_{B\nu]} \e^{AB} \right)\label{trasf:HM} \displaybreak[0] \\
\delta A_\mu^\L &= 2L^\L \ol\psi^A_\mu \epsilon^B \epsilon_{AB} + 2\ol
L^\L \ol\psi_{A\mu} \epsilon_B
\epsilon^{AB} + \ii f^\L_i \ol \l^{i A}\g_\mu \epsilon^B   \e_{AB}
  + \nn \\
& \qquad + \ii \ol  f^\tI_{\ib} \ol \l^{\ib}{}_A\g_\mu \epsilon_B
  \e^{AB}
\label{trasf:Flambda}\displaybreak[0]  \\
\delta A_{M\mu} &=2M_M \ol\psi^A_\mu \epsilon^B \epsilon_{AB} + 2\ol
M_M \ol\psi_{A\mu} \epsilon_B
\epsilon^{AB} + \ii g_{Mi} \ol \l^{i A}\g_\mu \epsilon^B   \e_{AB}
  + \nn \\
& \qquad \ii \ol  g_{M\ib} \ol \l^{\ib}{}_A\g_\mu \epsilon_B   \e^{AB}\label{trasf:FM} \displaybreak[0] \\
\delta z^i &=  \ol \epsilon_{A} \l^{i A} \label{trasf:zi} \displaybreak[0] \\
\d z^{\ib} &= \ol \epsilon^A \l^{\ib}{}_A
\label{trasf:zibar} \displaybreak[0] \\
\delta\psi_{A\mu} &= D_\mu \epsilon_A +   \e_{AB} T^-_{\mu\nu} \g^\nu \epsilon^B  +
 h_{\mu} \epsilon_A  + \ii S_{AB} \g_\mu \epsilon^B
 +\nonumber\\
& \qquad +\frac \ii 2 \epsilon_A \sx(Q_m  \ol  \psi_{B\mu} \lambda^{mB} +Q_\mb
 \ol  \psi_\mu^B \lambda^{\mb}_B\dx)-\frac \ii 2 \psi_{A|\mu} \sx(Q_m  \ol  \epsilon_{B} \lambda^{mB} +Q_\mb
 \ol  \epsilon^B \lambda^{\mb}_B\dx)\label{trasf:rhol}\displaybreak[0] \\
\delta\psi^A_{\mu} &= D_\mu \epsilon^A +   \e^{AB} T^+_{\mu\nu} \g^\nu \epsilon_B  -
 h_{\mu} \epsilon^A  + \ii \ol S^{AB} \g_\mu \epsilon_B
 + \nonumber\\
& \qquad -\frac \ii 2 \epsilon_A \sx(Q_m  \ol  \psi_{B\mu} \lambda^{mB} +Q_\mb
 \ol  \psi_\mu^B \lambda^{\mb}_B\dx)+\frac \ii 2 \psi_{A|\mu} \sx(Q_m  \ol  \epsilon_{B} \lambda^{mB} +Q_\mb
 \ol  \epsilon^B \lambda^{\mb}_B\dx) \label{trasf:rhor}\displaybreak[0] \\
 \d \l^{i A} &=  \ii D_\mu z^i \g^\mu \epsilon^A + G^{i-}_{\mu\nu}
 \g^{\mu\nu} \e^{AB}
\epsilon_B + W^{i AB} \epsilon_B
\label{trasf:lambdal} \displaybreak[0] \\
\d \l^{\ib}{}_A &=   \ii D_\mu \ol  z^{\ib} \g^\mu \epsilon_A + G^{\ib
  +}_{\mu\nu} \g^{\mu\nu} \e_{AB}
\epsilon^B + \ol {W}^{\ib}{}_{AB} \epsilon^B \label{trasf:lambdar} \displaybreak[0]
\end{align}

\subsection{The scalar potential}
To retrieve the equations of motion and in particular the scalar potential from the solution of Bianchi
identities would be a straightforward but very cumbersome computation. They can be derived in a much easier
way from the Lagrangian, which is presently under construction. We recall that in our complex formalism  the
dualization equations relating the antisymmetric tensors to the imaginary part of the scalars in the
vector-tensor sector are a consequence of the Bianchi identities, and therefore they are valid only
on-shell. Hence, prior to the explicit solution of the dualization equations, the scalar potential has
formally exactly the same structure as in standard $N=2$ theory, and it can be computed from the fermionic
shifts appearing in the supersymmetry transformation laws of the fermions \eq{trasf:rhol} -
\eq{trasf:lambdar}, namely:
\begin{equation}
V \delta^C_A=-12 \ol S^{BC} S_{AB} + g_{i\ol \jmath}W^{iBC} \ol W^{\ol \jmath}_{BA}\,.
\end{equation}

 The
fermion shifts can be read from the set of constraints on the gauge sector \eq{km} - \eq{constrgaugelast}.
 In the absence of hypermultiplets (and leaving aside possible Fayet--Iliopoulos terms)
they are:
\begin{align}
W^{i [AB]} &= \epsilon^{AB}\, \left(k^i_\L \ol  L^\L -k^{iM} \ol M_M\right)\\
S_{AB}&=0\,;\quad  W^{i (AB)} =0\,.
\end{align}
The above equations show that the only contributions to the scalar potential coming from the vector-tensor
sector are singlets of $SU(2)$ that belong to the $U(1)$ part of the R-symmetry. Note that the contribution
from the vector-tensor sector in $W^{i[AB]}$, in the absence of gauged isometries in the $G_0$ directions (that is if $k^m_\L = \delta^M_\Lambda k^m_M$), is symplectic invariant (recalling that $k^{iM}= \delta^i_m
k^{mM}$).

In the standard $N=2$ supergravity, the $SU(2)$ R-symmetry contribution to the scalar potential comes from
the hypermultiplet sector, where the $SU(2)$ is explicitly realized being part of the holonomy group of the
quaternionic manifold. The $SU(2)$  R-symmetry remains manifest when one considers the $N=2$, $D=4$
supergravity coupled to scalar-tensor multiplets \cite{Dall'Agata:2003yr,D'Auria:2004yi}\footnote{This is
obvious, since the scalar-tensor multiplet can be thought as resulting from dualization of (a subset of) the
scalars in the hypermultiplet sector.}. The relevant point is that the contribution of that sector to the
scalar potential is symplectic invariant.

We thus arrive at the important conclusion that $D=4$, $N=2$ supergravity coupled to scalar-tensor and
vector-tensor multiplets only, has a symplectic invariant scalar potential {\footnote{Assuming that there
are no gauged isometries in the graviphoton direction.}}.

We notice however that the study of the minima of the scalar potential requires an explicit form of the
geometry of the $\sigma$-model. This form has been computed explicitly in Appendix \ref{dualization} by
explicit dualization of the K\"ahler $\sigma$-model of a subsector of special geometry. Such dualization,
however, is a consequence of Bianchi identities and therefore it is valid only on-shell. This seems to be a
peculiarity of the formulation of $N=2$ supergravity coupled to vector-tensor multiplets.  In absence of a
factorization of the two $\sigma$-models of vector and vector-tensor multiplets, the explicit dualization
gives an on-shell geometry whose metric is not hermitian, even in the subsector pertaining to undualized
vector multiplets, as it is apparent from Appendix \ref{dualization}.

\section{Some comments on $D=5$, $N=2$ supergravity} \label{susy5}

The rich gauge structure underlying the bosonic sector of all FDA's can be in particular applied also to the
$N=2$, $D=5$ case. When this theory is constructed in the framework of FDA's with the rheonomic approach,
the final outcome is completely equivalent to the existing formulation in the literature \cite{
Gunaydin:1983rk,Gunaydin:1983bi,Gunaydin:1984pf,Gunaydin:1984ak,Gunaydin:1984nt,Sierra:1985ax,Lukas:1998yy,Lukas:1998tt,Gunaydin:1999zx,Gunaydin:2000xk,Gunaydin:2000ph,Gunaydin:2003yx,Ceresole:2000jd,Bergshoeff:2004kh,Gunaydin:2005bf}.
However, there are some features of the theory which are best appreciated in the light of the FDA approach,
on which we would like to comment. Such features concern in particular the Higgs mechanism, since in
our approach the 2-forms we start from are massless from the beginning. Let us then  discuss how the results
in the existing literature may be retrieved by starting with massless tensors\footnote{The explicit
construction of the theory within the present approach has been in fact given in \cite{Andrianopoli:2007xp}
(unpublished).}.

Actually, in our approach the mechanism for which the 2-forms become massive is left to the dynamics of the
Lagrangian (or alternatively, in the supersymmetric case, also of the supersymmetric Bianchi identities).
This is implemented via the Higgs mechanism. Even if  this procedure is very well
understood at the bosonic level, to implement it within a supersymmetric theory in $D=5$ is a non trivial task. This
is due to the fact that the supersymmetry constraints require the vectors $A^M$ giving mass to the tensors
(in the notations of section \ref{generalities}) to be related to the tensors themselves in a non local way,
involving Hodge-duality. This relation is codified in the so-called {\em ``self-duality-in-odd-dimensions"}
condition to which all the tensor fields in odd-dimensional supergravity theories have to comply
\cite{Townsend:1983xs}:
\begin{equation}
m^{MN}H_{N|abc} \propto \e_{abcde} F^{M|de}.\label{selfodd}
\end{equation}
In particular, for the five dimensional case the tensors are further required to be in even number.

Actually, in the approach currently adopted in the literature\break \cite{
Gunaydin:1983rk,Gunaydin:1983bi,Gunaydin:1984pf,Gunaydin:1984ak,Gunaydin:1984nt,Sierra:1985ax,Lukas:1998yy,Lukas:1998tt,Gunaydin:1999zx,Gunaydin:2000xk,Gunaydin:2000ph,Gunaydin:2003yx,Ceresole:2000jd,Bergshoeff:2004kh,Gunaydin:2005bf},
the tensors $B_M$ in the tensor multiplets are taken to be massive (and constrained to satisfy \eq{selfodd})
from the very beginning, without any tensor-gauge freedom.

To implement the Higgs mechanism on 2-forms at the supersymmetric level one could think of directly
supersymmetrizing the FDA \eq{fda}, and try to give mass to the whole tensor multiplets by coupling them to
$n_T$ extra abelian vector multiplets added to the theory:
\begin{equation}
(A^M_\mu, \chi^{M A},\phi^M),\label{false}
\end{equation}
 where the vectors $A^M$ and the
tensors $B_M$ admit the couplings and gauge invariance as in \eq{fda} and \eq{gaugefin}. If this would be
the case, in  the interacting theory the fields in the extra vector multiplets would couple to the tensor
multiplets and one would end up with $n_T$ {\em long} massive multiplets. We found,  however, from explicit
calculation  that this is not the case, since  supersymmetry transformations never relate the tensors $B_M$
to the spinors $\chi^{MA}$ nor to the scalars $\phi^M$ in \eq{false}. Then the only way compatible with
supersymmetry to couple  $N=2$ supergravity with $n_T$ massive tensors involves {\em
  short BPS} tensor multiplets
$$(B_{M|\mu\nu}, \lambda^{MA},\varphi^M)$$
where the massive tensors $B_M$ (that are complex, and hence in even number, because of CPT invariance of
the BPS multiplet) have to satisfy \eq{selfodd}. This is evident for the models having a six dimensional
uplift, where the mass of the tensors is the  BPS central charge  gauged by the graviphoton $g_{\mu
  5}$ \cite{Andrianopoli:2004xu}.
To show this,
let us look  at the
subclass of models obtained by Scherk--Schwarz dimensional reduction
from six dimensions. Indeed, the six-dimensional Lorentz algebra
admits as irreducible representations self-dual tensors, satisfying
\begin{equation}
\partial_{[\hat\mu} B_{\hat\nu\hat\rho]M} = \frac{1}{6}
\epsilon_{\hat\mu \hat\nu \hat\rho \hat\sigma \hat\lambda \hat\tau}
\partial^{\hat\sigma} B^{\hat\lambda \hat\tau}_M\,,\qquad \hat\mu
,\hat\nu,\dots = 0,1,\dots,5 .\label{sd6}
\end{equation}
Since $N=2$ matter-coupled supergravity in six dimensions contains one
antiself-dual and $n_T$ self-dual tensors in the vector representation
of $SO(1,n_T)$, one can use the $SO(n_T)\subset SO(1,n_T)$ global
symmetry of the model to dimensionally reduce the theory on a circle
down to five dimensions \`a la Scherk--Schwarz
\cite{Andrianopoli:2004xu}, with S-S phase $m^{MN}= - m^{NM} \in
SO(n_T)$:
\begin{equation}
B_{\hat\mu\hat\nu M} (x,y_5)=\Bigl({\exp}[m y_5]\Bigr)_M^{\phantom{M}N} \sum_n
B^{(n)}_{\hat\mu\hat\nu N}(x)\exp\left[\frac{\ii n }{2\pi
    R}y_5\right]\,.\label{ss}
\end{equation}
Applying \eq{ss} to the self-duality relation \eq{sd6} for the zero-mode, we find
\begin{equation}
\partial_{[\mu} B_{\nu\rho]M} = \frac16
\epsilon_{\mu\nu\rho\sigma\lambda 5}\left(m_M{}^N B_N^{\sigma\lambda} + 2 F^{\sigma\lambda}_M\right)
\,,\qquad \mu=0,1,\dots,4\label{sd5}
\end{equation}
where $F_{\sigma\lambda N}\equiv\partial_{[\sigma} B_{\lambda] 5 N}$. Eq. \eq{sd5} expresses the
self-duality obeyed by the tensors in five dimensional supergravity. However, it also shows that the
field-strengths of the vectors $B_{\mu 5 N}$, that give mass to the tensors $B_{\mu\nu M} $ via the
anti-Higgs mechanism, are in fact the Hodge-dual of the tensors $B_{\mu\nu M} $ themselves. From our
analysis applied to $N=2$ supergravity in five dimensions, we find this  to be a general fact, not
necessarily related to theories admitting a six dimensional uplift: in each case, the massive tensor fields
belong to short representations of supersymmetry, and the dynamical interpretation of the mechanism  giving
mass to the tensors requires the coupling of the massless tensors to gauge vectors which are the Hodge-dual
of the tensors themselves.


\section*{Acknowledgments}
We thank M. A. Lledo and especially Mario Trigiante for discussions and comments. One of us (R.D'A.) is
grateful to B. de Wit for  a kind discussion of his results on vector-tensor couplings in rigid
supersymmetry before publication.

Work supported in part by the European Community's Human Potential Program under contract
MRTN-CT-2004-005104 `Constituents, fundamental forces and symmetries of the universe', in which L.A. and
R.D'A. are associated to Torino University and L.S. to Valencia University, and by  PRIN Program 2005 of
University and Research Italian Ministry. The work of L.S. is partially supported from the Spanish Ministry
of Education and Science (SB-2005-0137,FIS2005-02761) and EU FEDER funds.


\appendix

\section{Constraints on 3-form Bianchi identities}
\label{constraints3f}

\begin{align}
m^{\L M} k^\a_M &= 0 \label{mk} \displaybreak[0] \\
f_{[\L\S}{}^\D f_{\G]\D}{}^\Pi &= 2 m^{\Pi M} L_{\L\S\G M} \displaybreak[0]  \\
f_{\S\G}{}^\L m^{\G M} &= m^{\L N} T_{\S N}{}^M  \displaybreak[0] \\
T_{\L M}{}^N &= d_{\L\S M} m^{\S N} + T_{\L\a}{}^N k^\a_M  \displaybreak[0] \\
T_{[\L| M}{}^P T_{\S] P}{}^N &= \mez f_{\L\S}{}^\G T_{\G M}{}^N +
k^\a_M N_{\a\L\S}{}^N + 3 L_{\L\S\G M} m^{\G N}  \displaybreak[0] \\
T_{\L M}{}^{(N} m^{\L |P)} &= k^\a_M N_\a{}^{NP}  \displaybreak[0] \\
T_{\L M}{}^N k_N^\a &= k^\b_M T_{\L \b}{}^\a  \displaybreak[0] \\
T_{[\L |M}{}^N L_{\S\G\D] N} &= k^\a_M M_{\L\S\G\D\a} + \frac{3}{2}
f_{[\L\S}{}^\Omega L_{\G\D]\Omega M}  \displaybreak[0] \\
T_{[\L |M}{}^N d_{\G|\S] N} &= \hat{f}_{\L\G}{}^\Pi d_{\Pi|\S]M} +
\mez f_{\L\S}{}^\Pi d_{\G\Pi M} + k^\a_M L_{\L\S\G\a} - 3 L_{\L\S\G M}
 \displaybreak[0] \\
T_{\L\a}{}^\b &= T_{\L\a}{}^M k_M^\b  \displaybreak[0] \\
T_{[\L|\a}{}^\g T_{\S]\g}{}^\b &= \mez f_{\L\S}{}^\G T_{\G\a}{}^\b +
k_M^\b N_{\a\L\S}{}^M  \displaybreak[0] \\
T_{\L\a}{}^\b m^{\L M} &= -2 N_\a{}^{MN} k_N^\b \displaybreak[0]  \\
T_{\L\a}{}^\b T_{\S\b}{}^M &= \hat{f}_{\L\S}{}^\Pi T_{\Pi\a}{}^M +
T_{\L\a}{}^N T_{\S N}{}^M - 2 L_{\L\Pi\G\a} m^{\Pi M} + \nn \\
& \qquad + 2 N_{\a\L\S}{}^M + 2 N_\a{}^{MN} d_{\S\L N}  \displaybreak[0] \\
T_{[\L|\a}{}^\b {T'}_{\S]\b}{}^M &= \mez f_{\L\S}{}^\G {T'}_{\G\a}{}^M
- {T'}_{[\L|\a}{}^N T^{(H)}_{\S]N}{}^M - L_{\L\S\G\a} m^{\G M} -
N_{\a\L\S}{}^M  \displaybreak[0] \\
T_{[\L|\a}{}^\b L_{\S\G]\D\b} &= T_{\D\a}{}^M L_{\L\S\G M} -
f_{[\L\S}{}^\Pi L_{\G]\Pi\D\a} + L_{[\L\S|\Pi\a} \hat{f}_{\G]\D}{}^\Pi
+ \nn \\
& \qquad + 4 M_{\L\S\G\D\a} + N_{\a[\L\S}{}^M d_{\D|\G] M}  \displaybreak[0] \\
T_{[\L|\a}{}^\b M_{\S\G\D\Pi]\b} &= 2 M_{\Omega[\L\S\G\a}
f_{\D\Pi]}{}^\Omega + N_{\a[\L\S}{}^M L_{\G\D\Pi]M}  \displaybreak[0] \\
T_{[\L|\a}{}^\b N_{\b|\S\G]}{}^M &= - 4 M_{\L\S\G\D\a} m^{\D M} -
N_{\a[\L|\Pi}{}^M f_{\S\G]}{}^\Pi + N_{\a[\L\S}{}^N T_{\G] N}{}^M +
\nn \\
& \qquad + 2 N_\a{}^{MN} L_{\L\S\G N}  \displaybreak[0] \\
T_{\L\a}{}^\b N_\b{}^{MN} &= - 2 N_{\a\L\S}{}^{(M} m^{\S|N)} + 2
N_\a{}^{P(M} T_{\L P}{}^{N)}  \displaybreak[0] \\
T_{\L\a}{}^M m^{\L N} &= {T'}_{\L\a}{}^N m^{\L M} - 2 N_\a{}^{MN}  \displaybreak[0] \\
T_{(\L|\a}{}^M d_{\S)\G M} &= - {T'}_{\G\a}{}^M d_{(\L\S)M} - 2
L_{\G(\L\S)\a} + R_{\b\L\S} (T_{\G\a}{}^\b + T'_{\a\G}{}^M k_M^\b ) +\nn\\
& \qquad + 2 R_{\a\Pi(\L} \hat{f}_{\G|\S)}{}^\Pi \label{constrlast} \displaybreak[0]
\end{align}


\section{Solution of superspace Bianchi identities  for $D=4$, $N=2$
  vector-tensor supergravity}
\label{biss}

Differentiating eqs. \eq{curvature} - \eq{lambdai} we get the supersymmetric BI's. Using the definitions:
\begin{align}
 D F^\L &= \de F^\L + \hat{f}_{\S\G}{}^\L A^\S F^\G + m^{\L N}
\hat{T}_{\S N}{}^M A_M F^\S  \displaybreak[0] \\
D F_M &= \de F_M + \hat{T}_{\L M}{}^N \sx( A^\L F_N - A_N F^\L \dx)
 \displaybreak[0] \\
D H_M &= \de H_M + \hat{T}_{\L M}{}^N A^\L H_N  \displaybreak[0] \\
D L^\L &= \de L^\L + \hat{f}_{\S\G}{}^\L A^\S L^\G + m^{\L N}
\hat{T}_{\S N}{}^M A_M L^\S + \ii \cQ L^\L \displaybreak[0] \\
D M_M &= \de M_M + \hat{T}_{\L M}{}^N \sx( A^\L M_N - A_N L^\L \dx) +
\ii \cQ M_M  \displaybreak[0] \\
D P_M &= \de P_M + \hat{T}_{\L M}{}^N A^\L P_N \displaybreak[0]
\end{align}
for the covariant derivatives of gauge-covariant quantities, where the gauged $\rm{U}(1)$ connection $\cQ $
has non vanishing components only along the $dz^x$ and $d{\bar z}^{\bar x}$ differentials, the Bianchi
identities read:

\begin{align}
D \cR^a{}_b &= 0  \displaybreak[0] \\
D T^a &+ \cR^a{}_b V^b - \ii \ol \psi^A \g^a \rho_A + \ii \ol \psi_A
\g^a \rho^A = 0 \displaybreak[0]  \\
\na \rho_A &+ \qu \g_{ab} \cR^{ab} \psi_A - \imez K \psi_A = 0  \displaybreak[0] \\
\na \rho^A &+ \qu \g_{ab} \cR^{ab} \psi^A + \imez K \psi^A  = 0  \displaybreak[0] \\
D F^\L &= D L^\L \ol \psi^A \psi^B \e_{AB} + D \ol L^\L \ol \psi_A
\psi_B \e^{AB} - 2 L^\L \ol \psi^A \rho^B \e_{AB} + \nn \\
& \quad - 2 \ol L^\L \ol \psi_A \rho_B \e^{AB} + m^{\L M} \sx( H_M - \ii
P_M \ol \psi_A \g_a \psi^A V^a \dx)  \displaybreak[0] \\
D F_M &= D M_M \ol \psi^A \psi^B \e_{AB} + D \ol M_M \ol \psi_A
\psi_B \e^{AB} - 2 M_M \ol \psi^A \rho^B \e_{AB} + \nn \\
& \quad - 2 \ol M_M \ol \psi_A \rho_B \e^{AB}  \displaybreak[0] \\
D H_M &= \ii D P_M \ol \psi_A \g_a \psi^A V^a - \ii P_M \sx( \ol
\psi_A \g_a \rho^A + \ol \psi^A \g_a \rho_A \dx) V^a + \nn \\
& \quad + P_M \ol \psi_A \g_a \psi^A \ol \psi_B \g^a \psi^B +  \nn \\
& \quad +\Bigl[d_{\L\S M}\sx( F^\S - L^\S \ol \psi^A \psi^B \e_{AB} - \ol L^\S \ol \psi_A \psi_B \e^{AB}
\dx) + \nn \\
& \,\qquad +\hat{T}_{\L M}{}^N \sx( F_N - M_N \ol \psi^A \psi^B \e_{AB} - \ol M_N \ol \psi_A \psi_B \e^{AB}
\dx)\Bigr] \cdot
\nn \\
& \quad \cdot \sx( F^\L - L^\L \ol \psi^C \psi^D \e_{CD} - \ol L^\L \ol \psi_C \psi_D \e^{CD} \dx)
\displaybreak[0] \\
D^2 z^i &= k^i_\L \sx( F^\L - L^\L \ol \psi^A \psi^B \e_{AB} - \ol
L^\L \ol \psi_A \psi_B \e^{AB} \dx)+ \nn \\
& \quad -   k^{mN} \sx( F_N - M_N \ol \psi^A \psi^B \e_{AB} -\ol M_N
\ol \psi_A \psi_B \e^{AB} \dx) \displaybreak[0]  \\
 \na^2
\l^{iA} &= \qu \g_{ab} \cR^{ab} \l^{iA} + \imez K \l^{iA} +  R^i{}_j \l^{jA} \displaybreak[0]\\
\na^2 \l^{\ib}_A &= \qu \g_{ab} \cR^{ab} \l^{\ib}_A - \imez K \l^{\ib}_A +  R^\ib{}_\jb \l^{\jb}_A
\displaybreak[0]\
\end{align}
where:
\begin{align}
K&=\de\mathcal{Q}
\end{align}
is the gauged K\"ahler 2-form.

\subsection{Parametrization of the curvatures in $D=4$, $N=2$
  superspace} \label{param4d}
\begin{align}
\mathcal{T}^a &= 0 \label{par:tors} \displaybreak[0]\\
\rho_A &= \rho_{A ab} V^a V^b +  \e_{AB} T^-_{ab} \g^b \psi^B V^a +
 h_a \psi_A V^a + \nonumber\\
& \qquad +\frac \ii 2 \psi_A \sx( Q_m \ol  \psi_B \lambda^{mB} + Q_\mb \, \ol  \psi^B \lambda^{\mb}_B\dx) +
\ii S_{AB} \g_a \psi^B V^a
\label{par:rhol}\displaybreak[0]\\
\rho^A &= \rho^A_{ab} V^a V^b +  \e^{AB} T^+_{ab} \g^b \psi_B V^a - h_a \psi^A V^a +\nonumber\\
&\qquad -\frac \ii 2 \psi_A \sx( Q_m \ol  \psi_B \lambda^{mB} + Q_\mb \, \ol  \psi^B \lambda^{\mb}_B\dx) +
\ii \ol S^{AB} \g_a \psi_B V^a
\label{par:rhor}\displaybreak[0]\\
H_M &= \hat H_{M|abc} V^a V^b V^c +  h_{Mi} \ol \psi_A \g_{ab} \l^{iA} V^a V^b +  h_{M\ib} \ol \psi^A
\g_{ab} \l^{\ib}{}_A V^a
V^b  \label{par:HM}\displaybreak[0] \\
F^\tI &= \hat F^\tI_{ab} V^a V^b + \ii f^\tI_i \ol \psi^A \g_a \l^{i B} \e_{AB} V^a + \ii \ol  f^\tI_{\ib}
\ol \psi_A \g_a \l^{\ib}{}_B \e^{AB} V^a
\label{par:Flambda} \displaybreak[0]\\
F_M &= \hat F_{M ab} V^a V^b + \ii g_{Mi} \ol \psi^A \g_a \l^{i B} \e_{AB} V^a + \ii \ol  g_{M\ib} \ol
\psi_A \g_a \l^{\ib}{}_B
\e^{AB} V^a \label{par:FM}\displaybreak[0] \\
D z^i &= D_a z^i V^a + \ol \psi_A \l^{i A} \label{par:zi}\displaybreak[0] \\
D z^{\ib} &= D_a \ol z^{\ib} V^a + \ol \psi^A \l^{\ib}{}_A
\label{par:zibar}\displaybreak[0] \\
\na \l^{i A} &= \hat \na_a \l^{i A} V^a + \ii D_a z^i \g^a \psi^A + G^{i-}_{ab} \g^{ab} \e^{AB} \psi_B +
W^{i AB} \psi_B
\label{par:lambdal} \displaybreak[0]\\
\na \l^{\ib}{}_A &= \hat \na_a \l^{\ib}{}_A V^a + \ii D_a \ol z^{\ib} \g^a \psi_A + G^{\ib +}_{ab} \g^{ab}
\e_{AB} \psi^B + \ol {W}^{\ib}{}_{AB} \psi^B \label{par:lambdar}\displaybreak[0]
\end{align}

\section{The vector-tensor $\sigma$-model metric}
\label{dualization}
Let us start from the (ungauged) kinetic term of special geometry:
\begin{eqnarray}
\cL_{kin} &=& g_{i\jb} \partial_\mu z^i \partial^\mu \ol z^\jb =\nn\\
& =& g_{r\sb} \partial_\mu z^r \partial^\mu \ol z^\sb +g_{m\nb}
\partial_\mu z^m \partial^\mu \ol z^\nb +\nn\\
&&+ g_{r\nb}
\partial_\mu z^r \partial^\mu \ol z^\nb + g_{m\sb} \partial_\mu z^m
\partial^\mu \ol z^\sb
 \,.
\end{eqnarray}
where we have denoted by $r,s$ the indices of the scalar fields of the vector multiplets which will not
undergo the dualization. Decomposing the differentials into real and imaginary parts $dz^i = dx^i +\ii
dy^i$, we easily get:
\begin{align}
\cL_{kin} = D_{ij} \left( \partial_\mu x^i \partial^\mu x^j +
\partial_\mu y^i \partial^\mu y^j\right) + 2
\G_{ij}\partial_\mu x^i \partial^\mu y^j \label{lagr1} \,,
\end{align}
where
\begin{align}
D_{ij}=\frac 12 (g_{i\jb} + g_{\ib j}) \quad &;& \quad
 \Gamma_{ij}=-\frac{\ii}{2} (g_{i\jb} - g_{\ib j})\,.
\end{align}
To perform the dualization on the vector-tensor multiplet sector, we
introduce the Lagrange multiplier $Y^M_\mu$ by the substitution
$\partial_\mu y^m \Rightarrow
\partial_\mu y^M \equiv Y^M_\mu$
and add to the Lagrangian \eq{lagr1} the term $\frac 1{3!} Y^M_\mu
H_{M \nu\rho\sigma} \epsilon^{\mu\nu\rho\sigma}$. Varying then the new
Lagrangian with respect to $Y^M_\mu$ one obtains
\begin{align}
Y^M_\mu &= -\frac 12 D^{MN}\left(2\Gamma_{iN} \partial_\mu x^i +2D_{rN}\partial_\mu y^r +\frac 1{3!}
H_{N}^{ \nu\rho\sigma} \epsilon_{\mu\nu\rho\sigma} \right)\label{ymmu},
\end{align}
where $D^{MN} $ is the inverse matrix of $D_{MN}$. Substituting \eq{ymmu} in \eq{lagr1} one obtains the dual
Lagrangian, namely:
\begin{eqnarray}
\cL_{dual} &=& \D_{ij} \partial_\mu x^i \partial^\mu x^j  + \tilde
\Delta_{rs} \partial_\mu y^r \partial^\mu y^s + 2\tilde \G_{ir}
\partial_\mu x^i \partial^\mu y^r+\nn\\
&& - \frac 1{3!} D^{MN}H_{M\nu\rho\sigma}\left[\frac 14 H_N^{\nu\rho\sigma} + \left(\G_{iN}\partial_\mu x^i
+D_{rN}\partial_\mu y^r \right)\epsilon^{\mu\nu\rho\sigma}\right],
\end{eqnarray}
where
\begin{eqnarray}
\D_{ij}&=& D_{ij} +  \G_{iM}D^{MN} \G_{Nj}\\
\tilde\D_{rs}&=& D_{rs} -  D_{rM}D^{MN} D_{Ns}\\
\tilde\G_{ir}&=& D_{ir} -  \G_{iM}D^{MN} D_{Nr}\,.
\end{eqnarray}
give the $\sigma$-model metric after dualization.


\section{Some notations for the four dimensional theories}\label{conv}

We use throughout the paper a mostly minus space-time metric $\eta_{ab}$.

The $\g_5$ matrix is defined as
\begin{align}
  \g_5 &= - i \g_0 \g_1 \g_2 \g_3\,.
  \end{align}
 Given an (anti-)self-dual tensor $F^\pm$, the following relations are
true:
\begin{align}
  {}^* F^\pm &= \mp i F^\pm
\end{align}
Furthermore, we use the following Fierz identities among spinor 1-forms:
\begin{align}
  \psi_A \ol \psi_B &= \mez \ol \psi_B \psi_A - \frac{1}{8} \g_{ab} \ol
  \psi_B \g^{ab} \psi_A \\
  \psi_A \ol \psi^B &= \mez \g_a \ol \psi^B \g^a \psi_A\,.
\end{align}
In terms of irreducible spinor representations, the following 3-gravitino relations hold:
\begin{align}
  \psi_A \ol \psi_B \psi_C &= \qu \e_{BC} \sx( \Theta^+_A - {\rm i}
  \, \Theta^-_A \dx)\\
  \psi_A \ol \psi^B \psi^C &= \qu \e^{BC} \sx( \Theta^+_A + {\rm i}
  \, \Theta^-_A \dx)\\
  \g_{ab} \psi_A \ol \psi_B \g^{ab} \psi_C &= 2 \e_{A(B} \sx(
  \Theta^+_{C)} - {\rm i} \, \Theta^-_{C)} \dx)\\
  \g_{ab} \psi^A \ol \psi^B \g^{ab} \psi^C &= 2 \e^{A(B} \sx(
  \Theta^{+C)} - {\rm i} \, \Theta^{-C)} \dx)\\
  \g_a \psi^A \ol \psi_B \g^a \psi^B &= - \mez \e^{AC} \sx(
  \Theta^+_C + {\rm i} \, \Theta^-_C \dx)
\end{align}


\end{document}